\documentclass[12pt, a4paper]{article}

\usepackage{amsmath,amssymb,bm,epsfig,afterpage}
\usepackage{cite}
\usepackage{here}
\usepackage{color}
\usepackage{colortbl}
\usepackage{xcolor}
\usepackage{tabularx}
\usepackage{subcaption}
\usepackage{bbm}
\usepackage{graphicx}
\usepackage{listings}
\usepackage{longtable}
\usepackage[utf8]{inputenc}
\usepackage{cancel}
\usepackage{slashed}
\usepackage{enumerate}
\usepackage{comment}

\graphicspath{{figures/}} 

\makeatletter
    
    \@addtoreset{equation}{section}
  \makeatother

\allowdisplaybreaks[3]

\newcommand{\beq}{\begin{equation}}
\newcommand{\eeq}{\end{equation}}

\newcommand{\ov}{\overline}

\newcommand{\ka}{\kappa}
\newcommand{\Ups}{\Upsilon}

\newcommand{\VEV}[1]{\left\langle #1 \right\rangle}

\renewcommand{\arraystretch}{1.3}

\newcommand{\GeV}{\mathrm{GeV}}
\newcommand{\TeV}{\mathrm{TeV}}

\newcommand{\eps}{\epsilon}

\newcommand{\tm}{\widetilde{m}}
\newcommand{\Mcal}{\mathcal{M}}
\newcommand{\Lcal}{\mathcal{L}}

\newcommand{\Ucal}{\mathcal{U}}
\newcommand{\Ncal}{\mathcal{N}}
\newcommand{\Dcal}{\mathcal{D}}
\newcommand{\Ecal}{\mathcal{E}}

\newcommand{\ev}{\mathbf{e}}
\newcommand{\dv}{\mathbf{d}}

\newcommand{\BL}{{B\mathrm{-}L}}

\newcommand{\eff}{\mathrm{eff}}
\newcommand{\VL}{\mathrm{VL}}
\newcommand{\Ptb}{P_{\overline{3}}}

\newcommand{\ol}[1]{\overline{#1}}

\newcommand{\vev}[1]{\langle{#1}\rangle}
\newcommand{\abs}[1]{\left|{#1}\right|}
\newcommand{\order}[1]{\mathcal{O}\left({#1}\right)}
\newcommand{\br}[2]{\mathrm{BR} \left({#1}\to{#2}\right)}
\newcommand{\id}[1]{\mathbf{1}_{{#1}} }

{\newcommand{\rep}[1]{\mathbf{#1}}
\newcommand{\bra}[1]{\langle{#1}| }
\newcommand{\ket}[1]{|{#1}\rangle }

\newcommand{\wm}{\widetilde{m}} 
\newcommand{\tF}{\widetilde{F}} 

\newcommand{\dL}[1]{d_L^{\;{#1}}}
\newcommand{\dR}[1]{d_R^{\;{#1}}}
\newcommand{\eL}[1]{e_L^{\;{#1}}}
\newcommand{\eR}[1]{e_R^{\;{#1}}}
\newcommand{\dbL}[1]{\overline{d}_L^{\;{#1}}}
\newcommand{\dbR}[1]{\overline{d}_R^{\;{#1}}}
\newcommand{\ebL}[1]{\overline{e}_L^{\;{#1}}}
\newcommand{\ebR}[1]{\overline{e}_R^{\;{#1}}}

\newcommand{\Deladj}{\Delta_8}

\usepackage[colorlinks=true, linkcolor=blue, citecolor=blue,
urlcolor=blue]{hyperref}

\setcounter{footnote}{0}

\begin{document}

\begin{titlepage}

\begin{flushright}
 {\tt
P3H-22-008,\,TTP22-001,\,CTPU-PTC-22-01
}
\end{flushright}

\vspace{1.2cm}
\begin{center}
{\Large
{\bf
Importance of 
vector leptoquark-scalar box diagrams in Pati-Salam unification with vector-like families
}
}
\vskip 1.5cm

{\large{Syuhei Iguro$^{1,2}$, Junichiro Kawamura$^3$, Shohei Okawa$^4$, and Yuji Omura$^5$}}

\vskip 0.5cm

$^{\rm 1}${\it Institute for Theoretical Particle Physics (TTP), Karlsruhe Institute of Technology (KIT),
Engesserstra{\ss}e 7, 76131 Karlsruhe, Germany}\\[3pt]
$^{\rm 2}${\it Institute for Astroparticle Physics (IAP),
Karlsruhe Institute of Technology (KIT), 
Hermann-von-Helmholtz-Platz 1, 76344 Eggenstein-Leopoldshafen, Germany}\\[3pt]

{\it $^3$
Center for Theoretical Physics of the Universe, Institute for Basic Science (IBS),
Daejeon 34051, Korea
}\\[3pt]

{\it $^4$ 
Departament de F\'isica Qu\`antica i Astrof\'isica, Institut de Ci\`encies del Cosmos (ICCUB),
Universitat de Barcelona, Mart\'i i Franqu\`es 1, E-08028 Barcelona, Spain
}\\[3pt]

{\it $^5$
Department of Physics, Kindai University, Higashi-Osaka, Osaka 577-8502, Japan}\\[3pt]

\vskip 1.0cm

\begin{abstract}
\vspace{0.1cm}
We study lepton flavor violation (LFV) induced by one-loop box diagrams 
in Pati-Salam (PS) unification with vector-like families. 
The vector leptoquark (LQ) associated with the PS gauge symmetry breaking generally causes various LFV processes such as $K_L \to \mu e$ and $\mu\to e$ conversion at the tree-level, thereby driving its mass scale to be higher than PeV scale.
The vector-like families are introduced to suppress such tree-level LFV processes,
allowing the LQ  to have TeV scale mass.
In this paper, we point out that there are inevitable one-loop contributions to those LFV processes from the box diagrams mediated by both one LQ and one scalar field, 
even if the tree-level contributions are suppressed. 
We consider a concrete model for demonstration, and show that the vector-like fermion masses have an upper bound for a given LQ mass when the one-loop induced processes are consistent with the experimental limits.
The vector-like fermion mass should be lighter than 3 TeV for 20 TeV LQ, 
if a combination of the couplings does not suppress $K_L \to \mu e$ decay.
Our findings would illustrate importance of the box diagrams 
involving both LQ and physical modes of symmetry breaking scalars in TeV scale vector LQ models.  
\end{abstract}
\vskip 1.0cm
KEYWORDS:\,{Pati-Salam,\,Vector leptoquark,\,Box diagrams,\,Lepton flavor violation}
\end{center}
\end{titlepage}

\tableofcontents
\clearpage

\section{Introduction}

The Pati-Salam (PS) unification~\cite{Pati:1974yy} is a compelling new physics scenario, 
given an ambitious motivation as a potential pathway toward grand unified theories.
In recent years, 
with growing interests in empirical hints for new physics found in flavor violating observables, significant attention has been paid to PS models and their variants with TeV scale symmetry breaking.
Indeed, a vector leptoquark (LQ), appearing as a result of the PS symmetry breaking, 
has the same quantum number as a  well-studied single mediator solution to the $R_{K^{(*)}}$~\cite{Aaij:2013qta,Aaij:2017vbb,Aaij:2014ora,Aaij:2019wad,Aaij:2015oid,Aaij:2020nrf,LHCb:2020gog,LHCb:2021lvy,Isidori:2020acz} and 
$R_{D^{(*)}}$~\cite{Lees:2012xj,Lees:2013uzd,Huschle:2015rga,Sato:2016svk,Hirose:2016wfn,Abdesselam:2019dgh,Aaij:2015yra,Aaij:2017deq,Iguro:2020cpg,Aoki:2021kgd} anomalies. 
Furthermore, such a particle could also account for the muon $g-2$ anomaly~\cite{Muong-2:2006rrc,Muong-2:2021ojo,Aoyama:2020ynm,Queiroz:2014zfa,Biggio:2016wyy}. 

In the basic construction of the PS unification, there are two known difficulties in lowering the PS symmetry breaking scale. 
Firstly, PS models partly unify quarks and leptons and hence predict the common mass matrices to them, which obviously contradicts the observed fermion spectra. 
Secondly, the vector LQ associated with the symmetry breaking carries both baryon and lepton numbers and 
couples quarks and leptons flavor-dependently at the tree-level. 
These couplings easily induce various flavor violating processes that are surppressed 
in the Standard Model (SM)~\cite{Bilenky:1987ty}.
In the conventional setup, the $K_L \to \mu e$ decay brings the most severe limit and pushes the PS breaking scale to be heavier than 1 PeV~\cite{Hung:1981pd,Valencia:1994cj}.

In recent studies, mainly stimulated by the $B$ meson anomalies,
it has been shown that 
both problems can be resolved by introducing vector-like fermions mixed with the SM chiral fermions~\cite{Calibbi:2017qbu,Dolan:2020doe,Iguro:2021kdw}. 
When specific relations between PS conserving vector-like masses 
and those originated from PS breaking are imposed, 
the vector LQ does not couple to a SM lepton and quark simultaneously 
and thus the serious lepton flavor violating (LFV) processes are absent at the tree-level, thereby allowing the TeV scale LQ. 
This solution, however, 
relies on a 0.1\% level tuning between the PS conserving and breaking masses 
to be consistent with the strong flavor constraints~\cite{Iguro:2021kdw}. 
We emphasize that the cancellations at the tree-level 
are not ensured by any symmetry, 
and thus loop-corrections possibly induce flavor violating processes in these models.

Loop-induced flavor violations in new physics models with LQs have been studied
in connection with the $B$ anomalies, e.g. based on a PS model~\cite{Calibbi:2017qbu}, 4321 models~\cite{DiLuzio:2018zxy,Fuentes-Martin:2020hvc}, extra dimension model~\cite{Blanke:2018sro} and composite model~\cite{Marzocca:2018wcf}.\footnote{See also Ref.~\cite{Crivellin:2018yvo}.}
In these works,  it turns out that the $b \to c$ transition correlates with the $B_s$ mixing induced by box diagrams with two LQs or two scalars.
The $R_{D^{(*)}}$ anomaly can be explained consistently 
only when the Glashow-Iliopoulos-Maiani (GIM) like mechanism~\cite{Glashow:1970gm} works for those box diagrams.

In this paper, we study another class of box diagrams which involve one vector LQ and one scalar, and evaluate how such diagrams impact on the LFV processes. 
As a demonstration, 
we consider a simple PS model only with vector-like copies of the SM chiral fermions   
and an $SU(4)_C$ adjoint scalar field 
in addition to the three generations of chiral fermions\footnote{
Additional scalar fields are necessary to break the residual $SU(2)_R \times U(1)_{B-L}$ gauge symmetry
as well as generating neutrino masses and mixings via the see-saw mechanism.
We do not discuss its explicit realization since this would not affect the flavor violations discussed in this work. 
An explicit model with $(\ol{\rep{10}}, \rep{1}, \rep{3})$ is studied in Ref.~\cite{Iguro:2021kdw}. 
}.  
In this setup, 
we compute all box diagrams relevant to the flavor violating processes 
associated with down-type quarks and charged leptons. 
Although several types of the Wilson coefficients 
turn out to be cancelled due to the GIM-like mechanism, 
we shall find that such cancellation does not work for contributions 
originated from box diagrams involving both LQ and PS breaking scalar. 
These contributions, thus, induce the LFV processes which are severely constrained. 
We also observe that the unsuppressed pieces are proportional 
to the powers of the ratio of the vector-like fermion mass to the LQ mass.
This indicates that the vector-like fermion masses are constrained from above 
for a given LQ mass. 
Interestingly, the vector-like fermions have to reside at the TeV scale or below if the PS symmetry breaks down at the TeV scale, which is suggested by the anomalies of the $B$-meson decays.

This paper is organized as follows. 
In Sec.\,\ref{sec-setup}, we introduce our PS model and explain how the observed mass spectra are realized by introducing vector-like fermions in addition to chiral ones. 
We also schematically explain how the box diagrams involving both LQ and scalar 
can induce sizable LFV processes. 
In Sec.\,\ref{sec-flv}, the box-induced contributions to the semi-leptonic operators are evaluated and the non-vanishing coupling structures are identified. 
We perform numerical analysis of the one-loop contributions in Sec.\,\ref{sec-pheno} and discuss the impact on the PS model construction.
Section \ref{summary} is devoted to summary.

\section{Pati-Salam model with TeV-scale vector LQ}
\label{sec-setup}

\subsection{Minimal Pati-Salam model} 

\begin{table}[t]
\centering
\caption{ \label{tab-mat}
The matter content in the Pati-Salam model.
}
\begin{tabular}{c|cccc}\hline
fields   & spin  & ~~$SU(4)_C$~~ & ~~$SU(2)_L$~~  & ~~$SU(2)_R$~~      \\ \hline\hline
  $L$    & $1/2$ & $\rep{4}$         &$\rep{2}$    &   $\rep{1}$       \\
  $F_L$  & $1/2$ & $\rep{4}$         &$\rep{2}$    &   $\rep{1}$       \\
  $F_R$  & $1/2$ & $\rep{4}$         &$\rep{2}$    &   $\rep{1}$       \\ 
\hline 
  $R$    & $1/2$ & $\rep{4}$         &$\rep{1}$    &   $\rep{2}$       \\
  $f_R$  & $1/2$ & $\rep{4}$         &$\rep{1}$    &   $\rep{2}$       \\ 
  $f_L$  & $1/2$ & $\rep{4}$         &$\rep{1}$    &   $\rep{2}$       \\ 
\hline 
 $\Delta$ & $0$ & $\rep{15}$       &$\rep{1}$ &   $\rep{1}$          \\
 $\Phi$& $0$   &  $\rep{1}$        &$\rep{2}$ &   $\ol{\rep{2}}$                    \\
\hline
\end{tabular}
\end{table}

We consider a model with the PS gauge symmetry, $SU(4)_C \times SU(2)_L\times SU(2)_R$. 
The PS symmetry identifies the lepton number as a fourth color of $SU(4)_C$, and 
SM chiral quark and lepton fields in the same $SU(2)_L$ representation are unified into chiral fields, $L_i$ and $R_i$ ($i=1,2,3$), 
whose representations are respectively 
$(\rep{4}, \rep{2}, \rep{1})$ and $(\rep{4}, \rep{1}, \rep{2})$ under the PS symmetry.
The electroweak (EW) doublet Higgs fields are embedded in a bi-doublet field of $SU(2)_L\times SU(2)_R$, $\Phi$, whose representation is $(\rep{1}, \rep{2}, \rep{\ol{2}})$. 
With this minimal content, the general Yukawa interaction is given by  
\begin{align}
\label{eq-yukSM}
-\Lcal^\mathrm{min}_Y=  \ol{L}  Y_{1} \Phi R+\ol{L}  Y_{2} \eps^T  \Phi^* \eps R +h.c.,
\end{align}
where $\eps := i\sigma_2$ acts on the $SU(2)_L$ and $SU(2)_R$ indices 
and $Y_1$ and $Y_2$ are $3\times 3$ Yukawa matrices in the flavor space.
A massive vector LQ, $X^\mu$, appears as a result of the $SU(4)_C \to SU(3)_C \times U(1)_{B-L}$ breaking. 
The LQ interaction is in the form of 
\begin{align}
\label{eq-LQcoup}
\Lcal_X & = \frac{g_4}{\sqrt{2}} X_\mu \left( \ol{d}_L^i \gamma^\mu e_L^i + \ol{u}_L^i \gamma^\mu \nu_L^i + \ol{d}_R^i \gamma^\mu e_R^i + \ol{u}_R^i \gamma^\mu \nu_R^i \right) + h.c. ,
\end{align}
which is written in the flavor basis and hence diagonal and universal 
unless the $SU(4)_C$ breaking effects to the fermion masses are taken into account. 
The CKM mixing factors will appear in the left-handed fermion interactions once we move to the mass basis. 
The interaction terms in Eqs.~(\ref{eq-yukSM}) and (\ref{eq-LQcoup}) explicitly show two problems inherent in the minimal PS model. 
One is that the minimal Yukawa terms in Eq.\,(\ref{eq-yukSM}) predict the same mass matrices for quarks and leptons, 
i.e. $m_e^{ij} = m_d^{ij}$ and $m_u^{ij} = m_\nu^{ij}$. 
The other is that the LQ interaction in Eq.\,(\ref{eq-LQcoup}) mediates $\ol{d}_i d_j \to \ol{e}_i e_j$ processes at the tree-level and triggers the rapid LFV processes, e.g. $K_L \to \mu e$.

\subsection{Schematic picture of tree-level flavor violation}

These problems can be resolved by introducing $SU(4)_C$ charged scalar fields and vector-like fermions in addition to $L^i$ and $R^i$. 
Before proceeding to the general argument, we briefly demonstrate our basic idea to create the mass splitting with a particular focus on the down-type quarks and charged leptons. 
The up-type quark and neutrino masses can be realized independently of the discussion below.
For a demonstrative purpose, we only focus on a single generation of $L^i$ and add one vector-like copy $F_{L,R} =(\rep{4}, \rep{2}, \rep{1})$ and an $SU(4)_C$ adjoint scalar field $\Delta$~\cite{Calibbi:2017qbu,Iguro:2021kdw}. 
With this extended content, we can write down additional interaction terms,
\begin{align}
 -\Lcal_M = \ol{L} \left( m_L + \ka_L \Delta \right) F_R + \ol{F}_L \left(M_L + Y_L \Delta \right) F_R 
              + h.c. ,  
\end{align}
where $m_L$ and $M_L$ are the PS symmetric mass parameters.
Decomposing the fermions 
\begin{align}
L = 
 \begin{pmatrix}
  \nu_L &  e_L \\
  u_L  &  d_L \\
 \end{pmatrix}, 
\quad 
F_A = 
 \begin{pmatrix}
  N_A &  E_A \\
  U_A  &  D_A \\
 \end{pmatrix}, 
\end{align}
with $A=L,R$ and 
after the adjoint field develops the vacuum expectation value (VEV) $\VEV{\Delta} = {v_\Delta}/{(2\sqrt{3})} {\rm diag}(3,-1,-1,-1)$, 
the mass terms take the form of 
\begin{align}
\label{eq-Mtoy}
 -\Lcal_M & = \ol{d_L} \left( m_L - \frac{\ka_L v_\Delta}{2\sqrt{3}} \right) D_R  + \ol{D_L} \left(M_L - \frac{Y_L v_\Delta}{2\sqrt{3}} \right) D_R \notag\\
 & + \ol{e_L} \left( m_L + \frac{3\ka_L v_\Delta}{2\sqrt{3}} \right) E_L + \ol{E_L} \left(M_L + \frac{3Y_L v_\Delta}{2\sqrt{3}} \right) E_R
              + \cdots,
\end{align}
where we only show the down-type quarks and charged leptons explicitly.
When the cancellation conditions, 
\begin{align}
\label{eq-tuning}
m_L - \frac{\ka_L v_\Delta}{2\sqrt{3}} = 0 , \quad 
M_L + \frac{3Y_L v_\Delta}{2\sqrt{3}} = 0 ,
\end{align}
are imposed, 
it follows that only $(D_L, D_R)$ and $(e_L, E_R)$ have vector-like masses,  
\begin{align}
 -\Lcal_M & = 
   \left(M_L - \frac{Y_L v_\Delta}{2\sqrt{3}} \right) \ol{D_L} D_R 
 + \left( m_L + \frac{3\ka_L v_\Delta}{2\sqrt{3}} \right) \ol{e_L} E_R 
              + \cdots,   
\end{align}
whereas $d_L$ and $E_L$ remain massless which are identified as the SM fermions. 
These massless quark and lepton respectively originate in the different PS multiplets, $L$ and $F_L$.
This indicates that unlike the minimal model, they have different Yukawa couplings to the bi-doublet scalar $\Phi$, which in turn results in different mass matrices for the quark and lepton after the EW symmetry breaking.
The left-handed LQ couplings are now in the form of 
\begin{align}
\label{eq-Xint}
\Lcal_X & = \frac{g_4}{\sqrt{2}}  X_\mu (\ol{d_L} \gamma^\mu e_L  + \ol{D_L} \gamma^\mu E_L) + \cdots. 
\end{align}
Since $e_L$ and $D_L$ are heavy, the LQ coupling always involves heavy fermions, 
not mediating the LFV meson decays at the tree-level. 
It should be noted that the EW gauge interactions remain unchanged 
under this exchange of the $SU(2)_L$ doublets $e_L$ and $E_L$. 
The mass splitting of the right-handed quark and lepton 
can be realized by introducing a corresponding vector-like copy in an analogous way.

\subsection{Schematic picture of loop-level flavor violation}

\begin{figure}[t]
\centerline{\includegraphics[viewport=100 550 470 770, clip=true, scale=0.7]{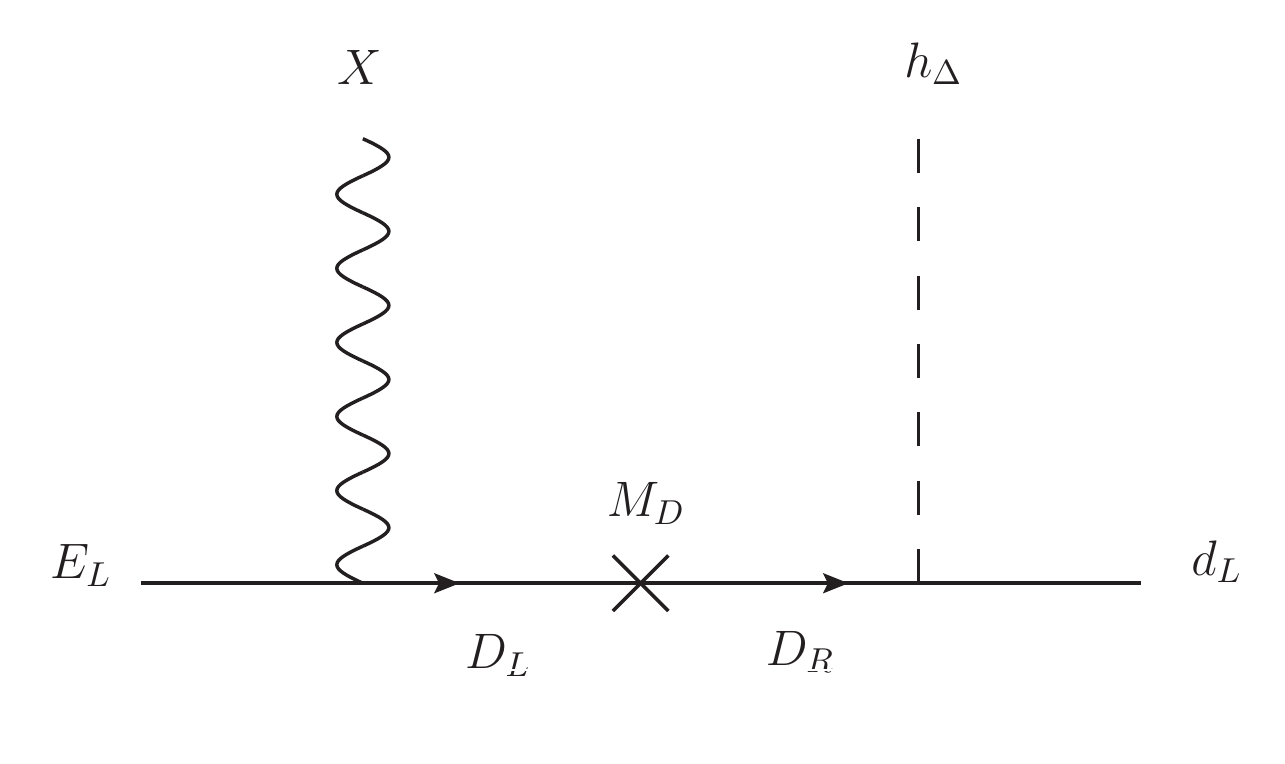}}
\caption{\label{X-hDelta} A schematic picture of regenerating the $d\to e$ transition via the interaction to the radial mode of the PS symmetry breaking scalar $h_\Delta$.
}
\end{figure}

Now we have seen that the mass splittings are generated by introducing the vector-like fermions. 
Following this method and adding more vector-like families, we can realize the observed mass spectra and well suppress the tree-level LFV processes in the three-generation case \cite{Iguro:2021kdw}.
On the other hand, this trick predicts a specific interaction structure that will revive the LFV processes via one-loop box diagrams.
To see this, we explicitly keep the radial mode of $\Delta$ by the replacement $v_\Delta \to v_\Delta + h_\Delta$ in Eq.\,(\ref{eq-Mtoy}). 
We then obtain Yukawa interactions of $h_\Delta$,
\begin{align}
\label{eq-hDeltaint}
 -\Lcal_M & = - \frac{\kappa_L}{2\sqrt{3}} \, h_\Delta \ol{d_L} D_R + \frac{3Y_L}{2\sqrt{3}} \, h_\Delta \ol{E_L} E_R
              + \cdots.
\end{align}
Although there is no mass term in the form of $\ol{d_L} D_R$ and $\ol{E_L} E_R$, 
the couplings of $h_\Delta \ol{d_L} D_R$ and $h_\Delta \ol{E_L} E_R$ are present. 
Thus, the $d \to e$ transition arises via the interaction to the radial mode $h_\Delta$, 
\begin{align}
\label{eq-dtoe}
\frac{g_4 \kappa_L}{2\sqrt{6} M_{D}} h_\Delta X_\mu \ol{d_L} \gamma^\mu E_L ,
\end{align}
where we assume the vector-like fermions are heavy and integrate them out (see also Fig.\,\ref{X-hDelta}). 
We note that $E_L$ is a SM-like lepton. 
Further integrating out the LQ and $h_\Delta$, we will obtain semi-leptonic operators that give rise to various LFV processes such as $K_L \to \mu e$ and $\mu\to e$ conversion.

One may wonder if other box diagrams involving two LQs or two $h_\Delta$ scalars can also induce the same semi-leptonic operators. 
This is true indeed, but as we will show in the rest of this paper, 
such contributions can be suppressed by the GIM-like mechanism. 
On the contrary, the LQ-scalar box contributions via the combination of Eqs.\,(\ref{eq-Xint}) and (\ref{eq-hDeltaint}) have no such suppression and thus are unavoidable. 
The evaluation of the LQ-scalar box diagrams is the main aim of this paper. 
In the following, we will study a realistic version of the PS extension highlighted above and 
perform analytical and numerical analyses to derive quantitative limits on the model. 

\subsection{Realistic model} 
In this section, we generalize the above discussion to accommodate the realistic quark and lepton masses. 
The matter contents of the model is summarized in Table\,\ref{tab-mat}.
We introduce $N_L$ vector-like $SU(2)_L$ doublet fermions $F_{L,R}$
and $N_R$ vector-like $SU(2)_L$ singlet fermions $f_{L,R}$ in addition to $L^i$, $R^i$ and $\Phi$.
Here $F_L$ and $f_R$ possess the same quantum numbers 
as those of $L$ and $R$, respectively, and $F_R$ and $f_L$ are their vector-like pairs. 
The relevant part of Lagrangian is given by 
\begin{align}
 -\Lcal_M = \ol{F}_L \left(M_L + Y_L \Delta \right) F_R + \ol{L} \left( m_L + \ka_L \Delta \right) F_R   
            + \ol{f}_L \left(M_R + Y_R \Delta \right) f_R + \ol{f}_L \left( m_R + \ka_R \Delta \right) R   
              + \cdots,  
\end{align}
where all couplings and masses are considered as matrices in the flavor space and 
the ellipsis represents the interactions between the fermions and the Higgs bi-doublet field $\Phi$.
The fermions are decomposed into quarks and leptons as 
\begin{align}
L = 
 \begin{pmatrix}
  \nu_L &  e_L \\
  u_L  &  d_L \\
 \end{pmatrix}, 
\quad 
R = 
 \begin{pmatrix}
  \nu_R &  e_R \\
  u_R  &  d_R \\
 \end{pmatrix},
\quad  
F_A = 
 \begin{pmatrix}
  N_A &  E_A \\
  U_A  &  D_A \\
 \end{pmatrix}, 
\quad 
f_A = 
 \begin{pmatrix}
  \Ncal_A &  \Ecal_A \\
  \Ucal_A  & \Dcal_A \\
 \end{pmatrix},
\end{align}
where the index $A=L,R$ represents the chirality. 
The scalar fields obtain non-vanishing VEVs, 
\begin{align}
 \vev{\Delta} =&\ \frac{v_\Delta}{2\sqrt{3}}
\begin{pmatrix}
 3 & 0 \\ 0 & -\id{3}
\end{pmatrix},
\quad
\vev{\Phi}
=v_H
\begin{pmatrix}
  \cos \beta & 0 \\ 0 & \sin \beta
\end{pmatrix} .
\end{align}

\subsection{Mass matrix}

We shall see the mass mixing of the fermions, which is directly linked to the coupling structure of the LQ and scalars. 
We mainly discuss the flavor violating processes among the down-type quarks and charged leptons in this paper, 
so that we only show the relevant terms.\footnote{See Ref.~\cite{Iguro:2021kdw} for the full detail of the mass mixing and diagonalization including the up-type quarks and neutrinos.}
Without loss of generality, 
the fermion mass terms are given by 
\begin{align}
\label{eq-MeMd}
- \Lcal_\mathrm{mass} =&\ \ol{\dv}_L\Mcal_d \dv_R + \ol{\ev}_L \Mcal_e \ev_R \\ \notag 
:=&\    
\begin{pmatrix}
 \ol{d}_L \\ \ol{D}_L \\ \ol{\Dcal}_L 
\end{pmatrix}^T 
\begin{pmatrix}
 m_{33} & m_{3R} & 0_{3\times N_L} \\ 
 m_{L3} & m_{LR} & D_{d_L} \\ 
 0_{N_R\times 3} & D_{d_R} & m_{RL}
\end{pmatrix} 
\begin{pmatrix}
 d_R \\ \Dcal_R \\ D_R 
\end{pmatrix}
+ 
\begin{pmatrix}
 \ol{e}_L \\ \ol{E}_L \\ \ol{\Ecal}_L 
\end{pmatrix}^T 
\begin{pmatrix}
 m_{33} & m_{3R} & M_{e E} \\ 
 m_{L3} & m_{LR} & M_{E E} \\ 
 M_{\Ecal e} & M_{\Ecal \Ecal} & m_{RL}
\end{pmatrix} 
\begin{pmatrix}
 e_R \\ \Ecal_R \\ E_R 
\end{pmatrix} 
\end{align}
where 
$D_{d_A}$ is an $N_A\times N_A$ diagonal matrix and $m_{\alpha \beta}$ $(\alpha,\beta = 3,L,R)$ 
is an $N_{\alpha}\times N_{\beta}$ mass matrix of $\order{v_H}$. Here, $N_\alpha = 3$ for $\alpha=3$. 
Note that the structure of $m_{\alpha \beta}$ is common to $\Mcal_d$ and $\Mcal_e$ due to the PS symmetry.
We define the mass basis for the fermions as 
\begin{align}
\hat{\dv}_{A} =
U_{d_A}^\dag  \dv_{A},
\quad
\hat{\ev}_{A} =
U_{e_A}^\dag  \ev_{A}.
\end{align}
The unitary matrices, $U_{f_A}$ ($f=e,d$), diagonalize the mass matrices as
\begin{align}
U_{f_L}^\dag 
\Mcal_f
U_{f_R} 
= \mathrm{diag}\left(m^f_1, m^f_2,m^f_3,\cdots, m^f_{N_L+N_R+3}  \right). 
\end{align}
The three lightest fermions correspond to the SM fermions and there are $N_L+N_R$ extra fermions.

We are interested in the approximate forms of the unitary matrices 
with $\eta := m_{\alpha\beta}/v_\Delta \ll 1$~\footnote{
If the $\order{v_H}$ entries are at most the bottom quark mass, 
we find $\eta \lesssim 10^{-3}$ for $v_\Delta \sim 4~\TeV$.}. 
Hence we only keep the leading contribution by neglecting ${\cal O}(\eta)$ corrections.
Assuming the elements of $D_{d_L}$ and $D_{d_R}$
are larger than the other elements,
the masses of the chiral three families $(d_L,~d_R)$ are mostly given by $m_{33} \sim {\cal O}(\eta)$. 
Then, $(d_L,~d_R)$ does not mix with the vector-like families
at the leading order in $\eta$ and can be regarded as SM-like.
On the other hand, the charged leptons $(e_L, e_R)$ do not correspond to the SM families due to the vector-like masses 
$M_{\Ecal e}$ and $M_{eE}$. 
The vector-like masses for the charged leptons can be, in general, decomposed as 
\begin{align}
 \begin{pmatrix}
  M_{eE} \\ M_{EE} 
 \end{pmatrix}
 =: 
 V_L 
\begin{pmatrix}
D_{e_L} \\  0_{3 \times N_L} 
\end{pmatrix}
 W_R^\dagger, 
\quad 
  \begin{pmatrix}
  M_{\Ecal e} & M_{\Ecal \Ecal} 
 \end{pmatrix}
 =: 
  W_L
\begin{pmatrix}
 0_{N_R\times 3} & D_{e_R}
\end{pmatrix}
V_R^\dagger,  
\end{align}
where $D_{e_A}$ is a diagonal $N_{A}\times N_A$ matrix with real positive entries, $V_A$ and $W_A$ are $(3+N_A)\times (3+N_A)$ 
and $N_{\ol{A}}\times N_{\ol{A}}$ unitary matrices and $\ol{A}$ denotes the opposite chirality of $A$ (i.e. $\ol{A} = R,L$ for $A=L,R$). 
Hence, the unitary matrices are approximately given by 
\begin{align}
 U_{d_L} =&\  
\begin{pmatrix}
 \id{3} & 0 & 0 \\ 
 0 & 0 & \id{N_L} \\
 0 & \id{N_R} & 0 \\   
\end{pmatrix}, 
\quad 
 U_{d_R} = 
\begin{pmatrix}
 \id{3} & 0 & 0 \\ 
 0 & \id{N_R} & 0 \\ 
 0 & 0 & \id{N_L} \\   
\end{pmatrix}, 
\\  \notag 
U_{e_L}= &\ 
\begin{pmatrix}
 V_L & 0_{(3+N_L) \times N_R} \\ 
 0_{N_R \times (3+N_L)} & W_L 
\end{pmatrix}
P,  
\quad 
U_{e_R}=  
\begin{pmatrix}
  0_{(3+N_R) \times N_L} & V_R \\ 
W_R &  0_{N_L \times (3+N_R)} 
\end{pmatrix}
P,
\end{align} 
with 
\begin{align}
P:= 
 \begin{pmatrix}
 0 & 0 & \id{N_L} \\
 \id{3}& 0 & 0 \\ 
 0 & \id{N_R} & 0 
\end{pmatrix}.
\end{align}
The matrix $V_L$ $(V_R)$ represents the mixing within the $SU(2)_L$ doublets (singlets) in the left-handed (right-handed) sectors, while the matrix $W_L$ $(W_R)$ is the mixing within the $SU(2)_L$ singlets (doublets) in the left-handed (right-handed) sector.
The vanishing blocks of $U_{e_A} P^{-1}$ reflect the fact that the $SU(2)_L$ doublet and singlet do not mix each other without the VEV of $\Phi$. 
These blocks have non-vanishing entries of ${\cal O}(\eta)$ in fact.
With the unitary matrices $U_{d_A}$ and $U_{e_A}$ in this form, the mass matrices are approximately diagonalized as
\begin{align}
U_{d_L}^\dag \Mcal_d U_{d_R} = 
\begin{pmatrix}
 m_{33} & m_{3R} & 0 \\ 
 0 & D_{d_R} & m_{RL} \\ 
 m_{L3} & m_{LR} & D_{d_L} 
\end{pmatrix}
=: D_d 
, 
\quad 
U_{e_L}^\dag \Mcal_e U_{e_R} = 
\begin{pmatrix}
 \wm_{33} & \wm_{3R} & 0 \\ 
 0 & D_{e_R} & \wm_{RL} \\ 
 \wm_{L3} & \wm_{LR} & D_{e_L} 
\end{pmatrix}
=: D_e 
, 
\end{align}
where 
\begin{align}
 V_L^\dag 
\begin{pmatrix}
 m_{33}& m_{3R} \\ m_{L3} & m_{LR}
\end{pmatrix}
V_R 
=: 
\begin{pmatrix}
 \wm_{L3}& \wm_{LR} \\ \wm_{33} & \wm_{3R} 
\end{pmatrix},
\quad 
W_L^\dag m_{RL} W_R =: \wm_{RL}.  
\end{align}
Neglecting the sub-dominant effects suppressed by $\eta$,  
the mass matrices of the SM down-type quarks and charged leptons are determined by $m_{33}$ and $\wm_{33}$, respectively.   
Their singular values should be consistent with the SM fermion masses. 
The other elements of $m_{\alpha\beta}$ and $\wm_{\alpha\beta}$ give only sub-dominant effects to the mixing matrices 
as far as the vector-like fermions are sufficiently heavier than the EW scale. 

\subsection{Couplings with the vector LQ and scalars} 

The $SU(4)_C$ gauge symmetry is broken by the VEV of the adjoint scalar $\Delta$. 
The massive gauge boson $X_\mu$ associated with the $SU(4)_C \to SU(3)_c \times U(1)_{B-L}$ breaking arises as a vector LQ.
If $\Delta$ is the only source of the $SU(4)_C$ symmetry breaking, the LQ mass,  $m_X$ is given by 
\begin{align}
\label{eq;LQmass}
m_X = \frac{2 }{\sqrt{3}} g_4 v_\Delta ,
\end{align}
where $g_4$ denotes the $SU(4)_C$ gauge coupling. 
This relation is modified if other scalars contribute to the $SU(4)_C$ breaking.

The LQ couplings are given by the gauge interactions of the LQ to the fermions
\begin{align}
\label{eq-Xcoup}
\Lcal_X =&\  \frac{g_4}{\sqrt{2}} X^\mu
                    \Bigl(\ol{\rep{d}}_L \gamma_\mu \rep{e}_L +
                     \ol{\rep{d}}_R \gamma_\mu  \rep{e}_R
                     \Bigr)  + h.c.   \\  \notag
           =&\ X^\mu 
      \Bigl(\ol{\hat{\rep{d}}}_L \hat{{g}}_{L}  \gamma_\mu  \hat{\rep{e}}_L  
            +   \ol{\hat{\rep{d}}}_R \hat{{g}}_{R} \gamma_\mu   \hat{\rep{e}}_R 
      \Bigr) + h.c.,
\end{align}
where the LQ couplings in the mass basis are expressed in terms of the fermion mixing matrices,
\begin{align}
\hat{g}_L = \frac{g_4}{\sqrt{2}} \Omega_L,
\quad 
\hat{g}_R = \frac{g_4}{\sqrt{2}} \Omega_R, 
\quad  
\mathrm{with}
\quad 
\Omega_L  :=  U_{d_L}^\dag U_{e_L}, 
\quad 
\Omega_R  := U_{d_R}^\dag U_{e_R}. 
\end{align}
For a practical purpose, it is useful to decompose the unitary matrices $V_{L,R}$ as
\begin{align}
 V_L =: 
\begin{pmatrix}
 V_{3L} & X_L \\ Y_L & V_{L3}  
\end{pmatrix}, 
\quad 
 V_R =: 
\begin{pmatrix}
 X_R & V_{3R} \\ V_{R3} & Y_R 
\end{pmatrix}, 
\end{align}
with which we find $\Omega_{L,R}$ to be 
\begin{align}
\label{eq-OmgLR}
 \Omega_L = 
\begin{pmatrix}
X_L & 0 & V_{3L} \\ 0 & W_L & 0 \\ V_{L3} & 0 & Y_L 
\end{pmatrix}, 
\quad 
 \Omega_R = 
\begin{pmatrix}
X_R & V_{3R} & 0 \\ V_{R3} & Y_R & 0 \\  0 & 0 & W_R \\ 
\end{pmatrix}. 
\end{align}
It should be noted that we neglected the ${\cal O}(\eta)$ contribution above. 
The $3\times 3$ matrix $X_A$ in $V_A$ represents the overlap of the SM charged leptons $\hat{e}^i_A$ with the first three PS multiplets $L^i$ and $R^i$ which the SM down-type quarks $\hat{d}^i_A$ mostly originate in. 
Thus, it follows from Eq.\,(\ref{eq-OmgLR}) that $X_A$ stands for the LQ couplings to two SM fermions. 
As a reminder, taking $X_A = 0$ corresponds to the exact cancellation in Eq.\,(\ref{eq-tuning}). 

After the $SU(4)_C$ breaking, the adjoint scalar $\Delta$ is decomposed into a singlet scalar $h_\Delta$ and an $SU(3)_c$ adjoint scalar $\Delta_8$:
\begin{align}
 \Delta =&\ \frac{1}{2\sqrt{3}} \left( v_\Delta + \frac{h_\Delta}{\sqrt{2}} \right)
\begin{pmatrix}
3&0  \\
0&-\rep{1}_3 \\
\end{pmatrix}
+
\begin{pmatrix}
 0 & 0 \\ 0 & \Deladj
\end{pmatrix}, 
\end{align}
where $\id{n}$ denotes an $n\times n$ identity matrix. 
The Yukawa couplings involving $h_\Delta$ and $\Deladj$ are given by
\begin{align}
- \mathcal{L}_{\Delta} =
 \sum_{\rep{f}=\rep{d},\rep{e}}
       \sqrt{\frac{{3}}{8}} h_\Delta Q_{\BL}^{\rep{f}} \ol{\rep{f}}_L  Y_\Delta \rep{f}_R
+\Deladj \ol{\rep{d}}_L  Y_\Delta \,  \rep{d}_R
+ h.c.,
\end{align}
where $Q_{\BL}^{\rep{d}}=-1/3$ and $Q_{\BL}^{\rep{e}}=+1$ are the $\BL$ charges. 
The Yukawa couplings in the mass basis are given by
\begin{align}
\hat{Y}_\Delta^f = \left(U^f_L\right)^\dagger Y_\Delta U^f_R ,  \quad f=d,e.
\end{align}
There is a close relation between the Yukawa coupling matrices in the gauge basis and the fermion mass matrices, 
\begin{align}
 Y_\Delta =
\frac{\sqrt{3}}{2 v_\Delta} \left(\Mcal_{e}-\Mcal_{d} \right).
\label{Yukawa-Delta}
\end{align}
Using this relation, 
the Yukawa matrices in the mass basis are expressed as 
\begin{align}
\label{eq-YDed}
 \hat{Y}_\Delta^e = \frac{\sqrt{3}}{2v_\Delta} 
            \left( D_e - \Omega_L^\dag D_d \Omega_R \right) + \order{\eta}.   
\quad 
\hat{Y}_\Delta^d = \frac{\sqrt{3}}{2v_\Delta} 
            \left(\Omega_L D_e \Omega_R^\dag - D_d \right) + \order{\eta}.  
\end{align}
The flavor violating couplings are thus induced via 
\begin{align}
\Omega_L^\dag D_d \Omega_R \sim&\  
 \begin{pmatrix}
  0 & 0 & V_{L3}^\dag D_{d_L} W_R \\ 
  W_L^\dag D_{d_R} V_{R3} & W_L^\dag D_{d_R} Y_R & 0  \\ 
 0 & 0 & Y_L^\dag D_{d_L} W_R 
 \end{pmatrix} 
+ \order{m_{\alpha\beta}},
\label{eq:Omega_LDR}
\end{align}

\begin{align}
\Omega_L D_e \Omega_R^\dag \sim&\  
\begin{pmatrix}
0 & 0 & V_{3L} D_{e_L} W_R^\dag \\ 
 W_L D_{e_R} V_{3R}^\dag & W_L D_{e_R} Y_R^\dag & 0 \\ 
 0 & 0 & Y_L D_{e_L} W_R^\dag 
\end{pmatrix}
+ \order{m_{\alpha\beta}} .
\label{eq:Omega_LER}
\end{align}
One can see that the tree-level couplings of $h_\Delta$ and $\Delta_8$ to the SM families are suppressed by $\eta$.

\section{Flavor violations from box diagrams} 
\label{sec-flv} 

In this section, we look at box contributions to the flavor violating processes, 
using the LQ and scalar couplings to the fermions derived in the previous section. 

\subsection{Tree-level constraints}

We first summarize
the constraints set by considering only the tree-level LQ exchange, which motivate us to impose a primary suppression condition.
It follows from Eq.\,\eqref{eq-OmgLR} that the LQ couplings to the SM fermions are given by $X_L$ and $X_R$.
With these couplings, the tree LQ exchange induces the semi-leptonic operators in the form of
\begin{equation}
\Lcal_{\rm eff} = \frac{g_4^2}{2m_X^2} \sum_{A,B=L,R} (X_A)^{ik} (X_B^\dagger)^{lj} (\ol{d}_A^i \gamma^\mu e_A^k) (\ol{e}_B^l \gamma_\mu d_B^j) ,
\end{equation}
leading to various LFV processes, e.g. $K_L \to \mu e$, $B_d \to \tau e$, $B_s \to \tau \mu$ and $\mu\to e$ conversion. 
We show in Appendix\,\ref{sec:tree_constraint} a rough estimate of experimental bounds on $X_L$ and $X_R$, assuming $m_X =5$\,TeV and $g_4=1$. 
This set of bounds suggests that if all elements of $X_A$ have comparable size, 
each matrix element should satisfy 
\begin{equation}
\left |(X_L)^{ij} \right |, \, \left |(X_R)^{ij}\right | \lesssim \mathcal{O}(10^{-3}) .
\end{equation}
This limit motivates us to impose a condition for the suppressed flavor violation:
\begin{itemize}
 \item [(i)]{$X_L = X_R = 0$} .
\end{itemize}
Given this condition, the unitarity of $V_A$ requires that $N_{A} \ge 3$ and 
\begin{align}
\label{eq-unitV3A}
 V_{3A} V_{3A}^\dag \simeq V_{A3}^\dag V_{A3} \simeq \id{3} .
\end{align}
Note that the exchange of $h_\Delta$ and $\Delta_8$ does not cause any flavor violation at the tree-level,
since their Yukawa couplings always contain the heavy fermions as seen in Eqs.\,\eqref{eq-YDed}, (\ref{eq:Omega_LDR}) and (\ref{eq:Omega_LER}). 
Therefore, once the condition (i) is imposed, the model is free from the flavor violation at the tree-level. 

\subsection{Box contributions} 

\begin{figure}[t]
\vspace{-4cm}
\centerline{\includegraphics[scale=0.4]{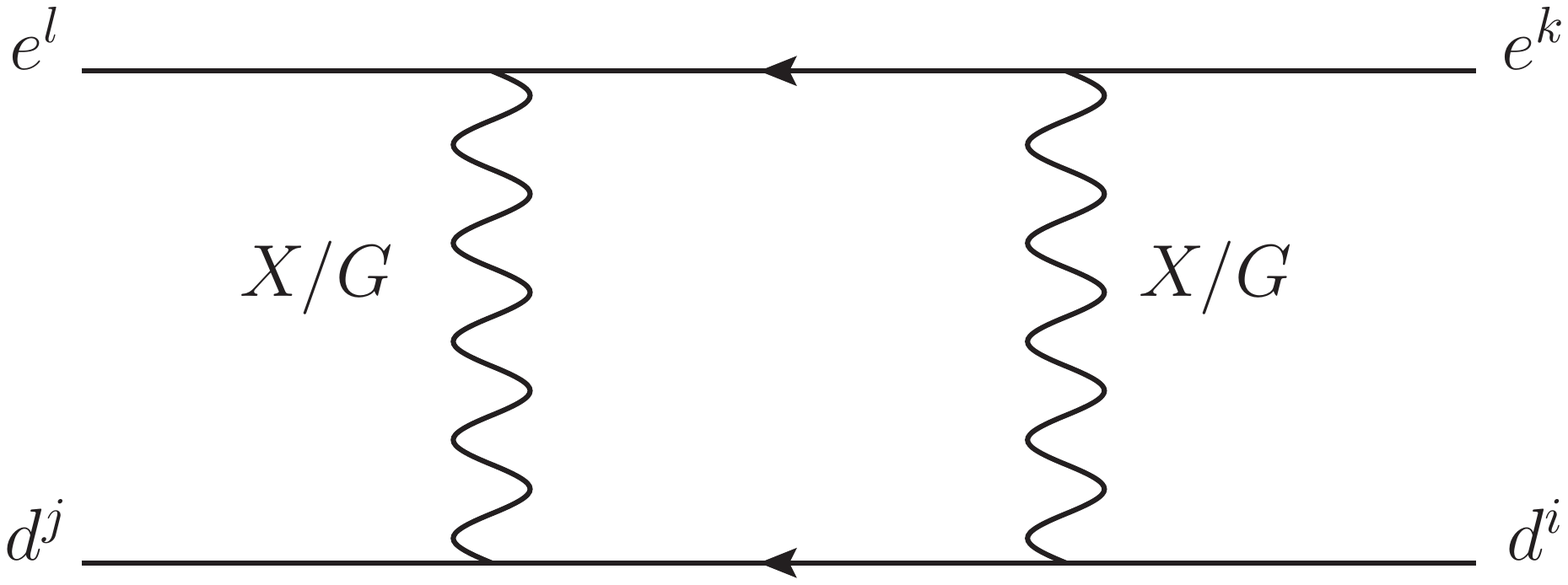}}
\vspace{-4cm}
\caption{\label{Box-XX} 
The box diagram involving LQ and Goldstone boson.
}
\end{figure}
\begin{figure}[t]
\centerline{
\includegraphics[scale=0.12]{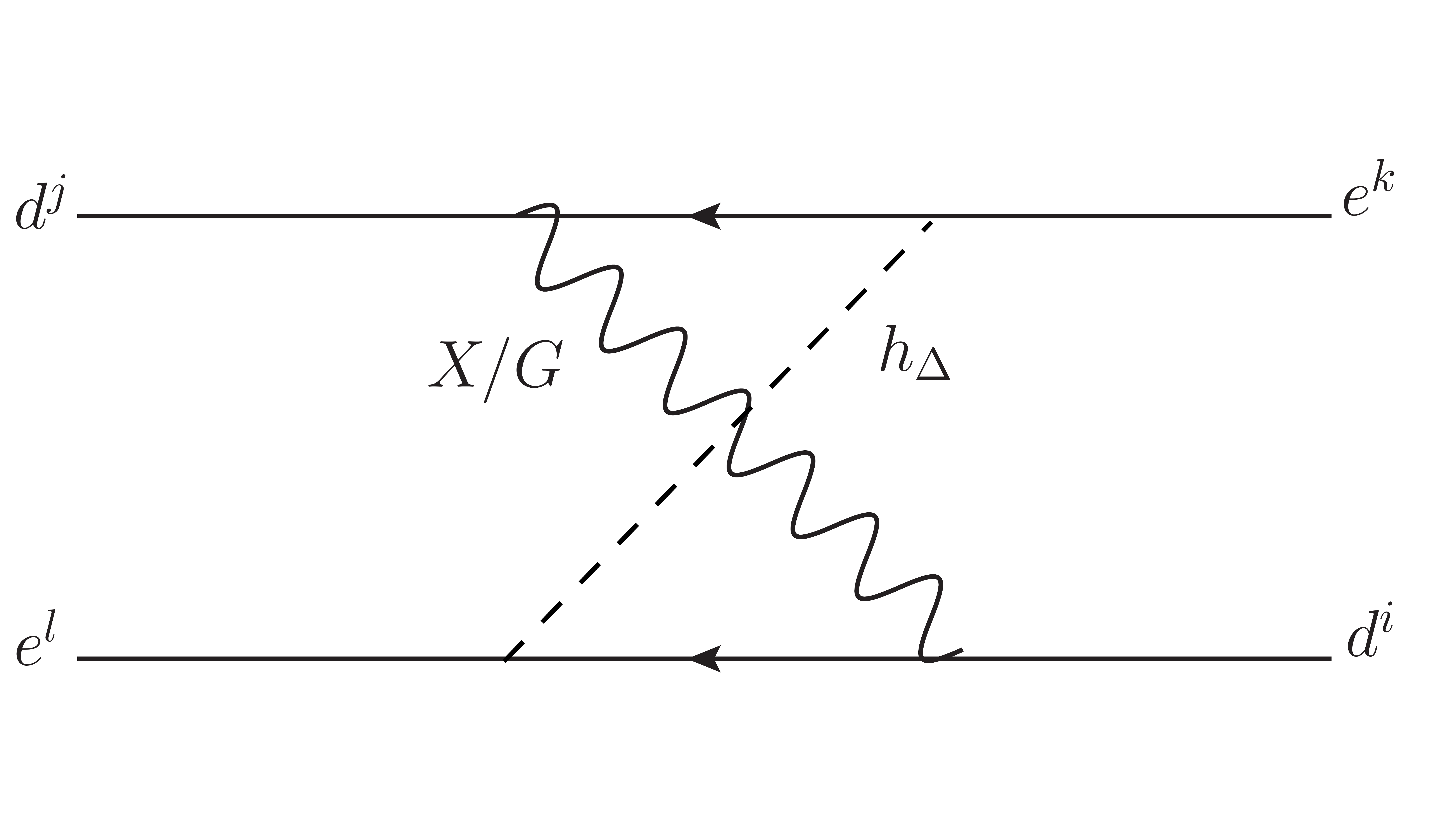}~~~
\includegraphics[scale=0.12]{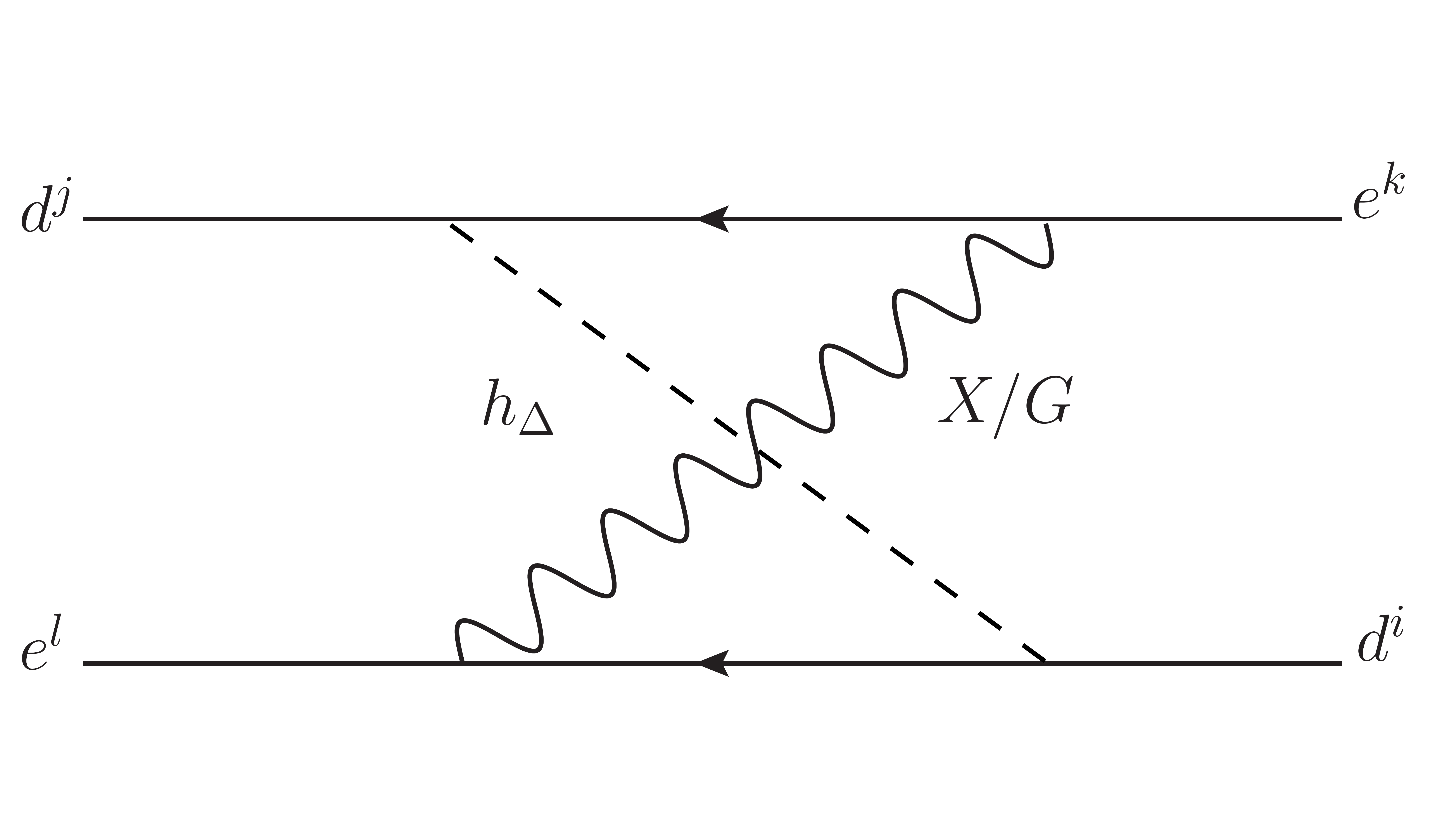}
}
\centerline{
\includegraphics[scale=0.12]{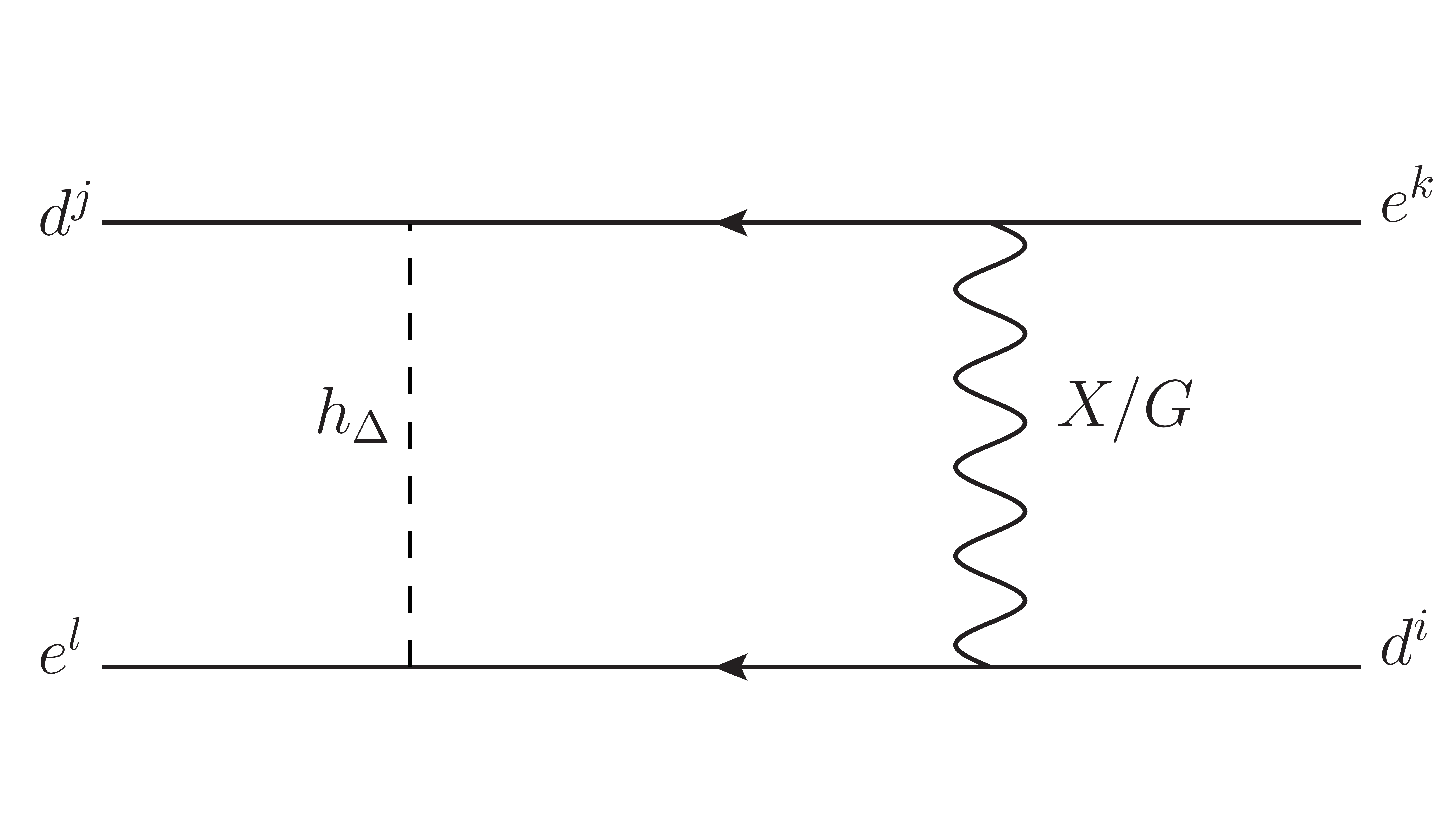}~~~
\includegraphics[scale=0.12]{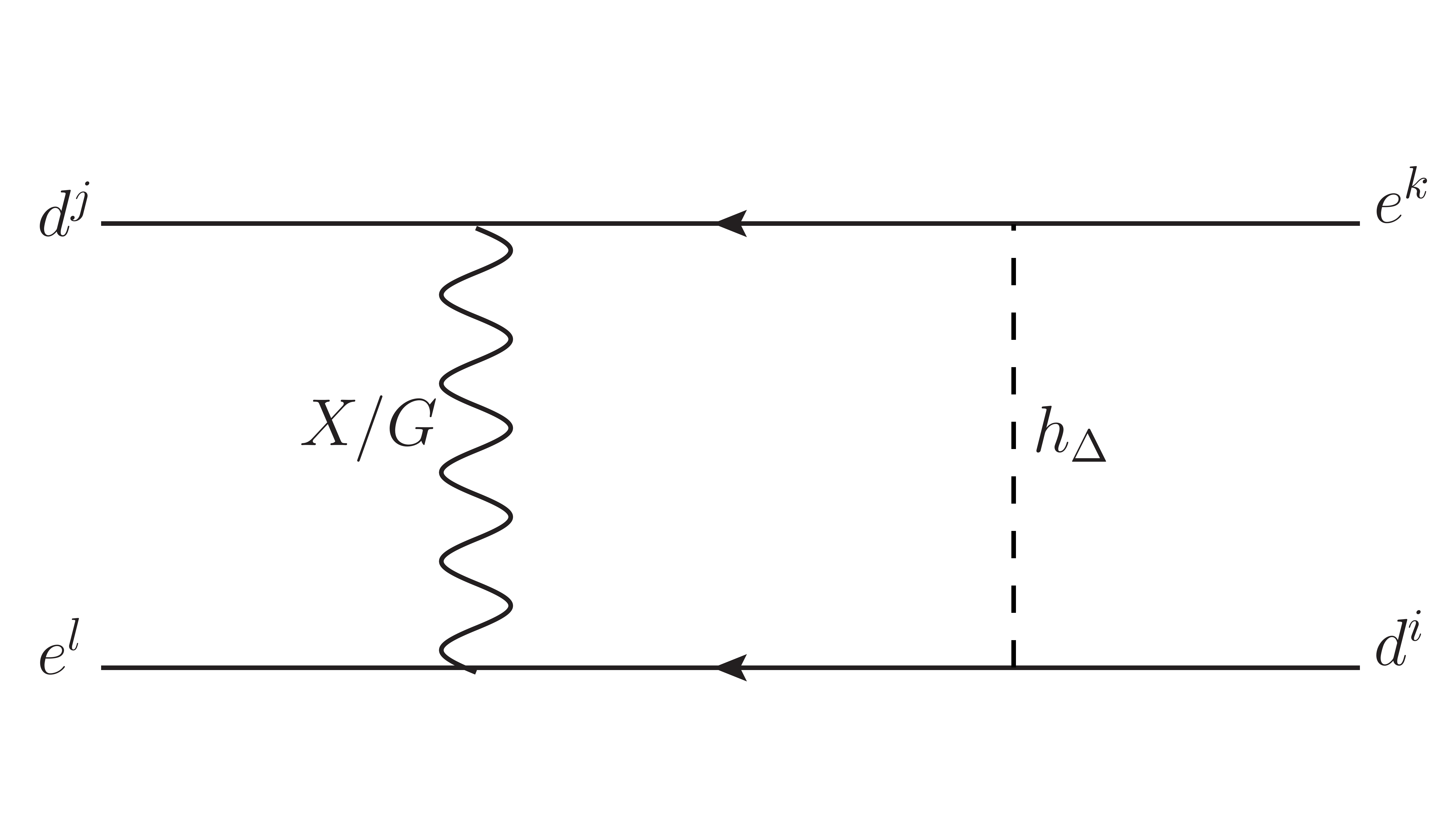}
}
\centerline{
\includegraphics[scale=0.12]{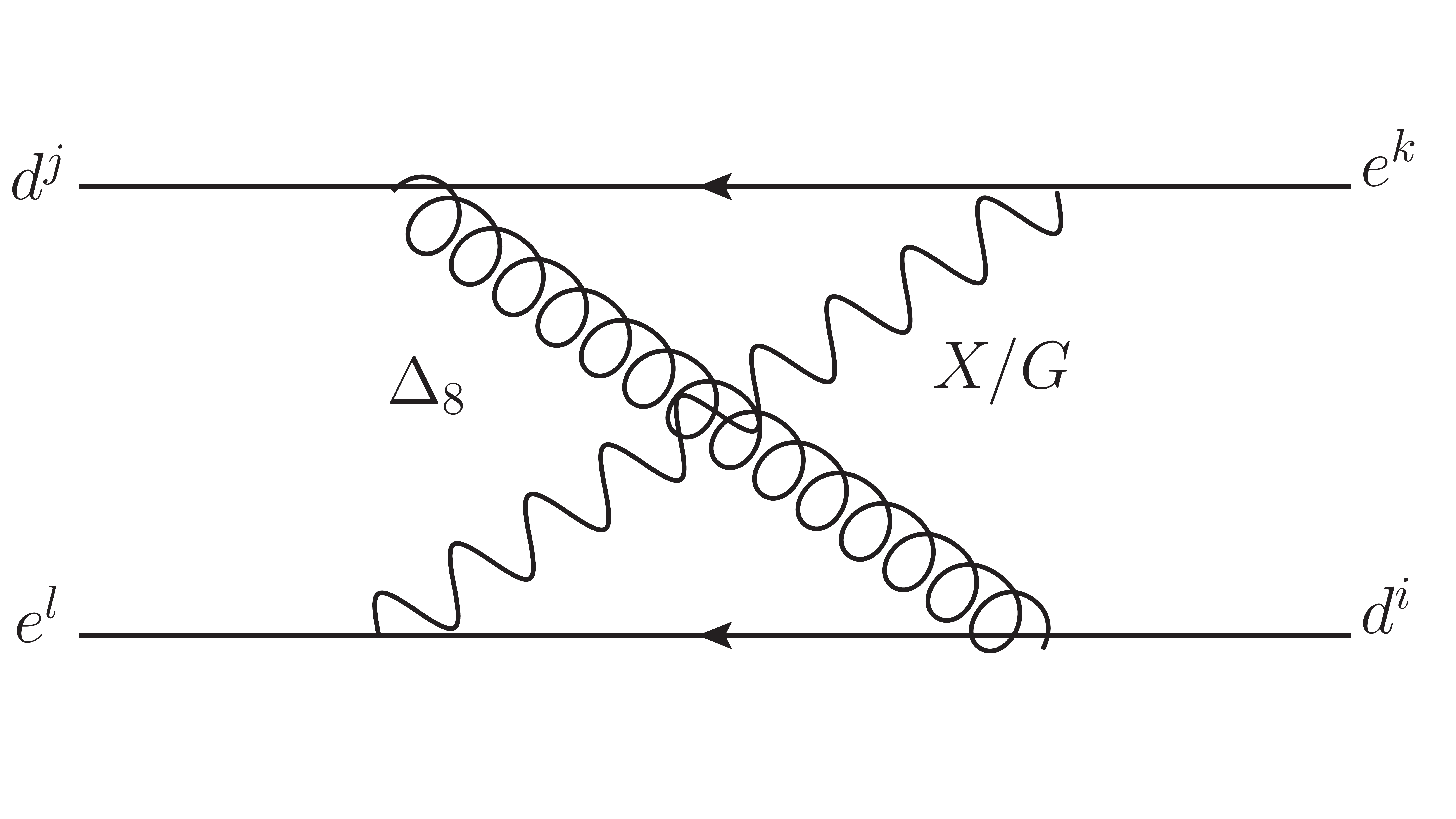}
}
\caption{\label{Box-Xh56} The box diagrams involving scalars and LQ/Goldstone.}
\end{figure}

\begin{figure}[t]
\centerline{
\includegraphics[scale=0.12]{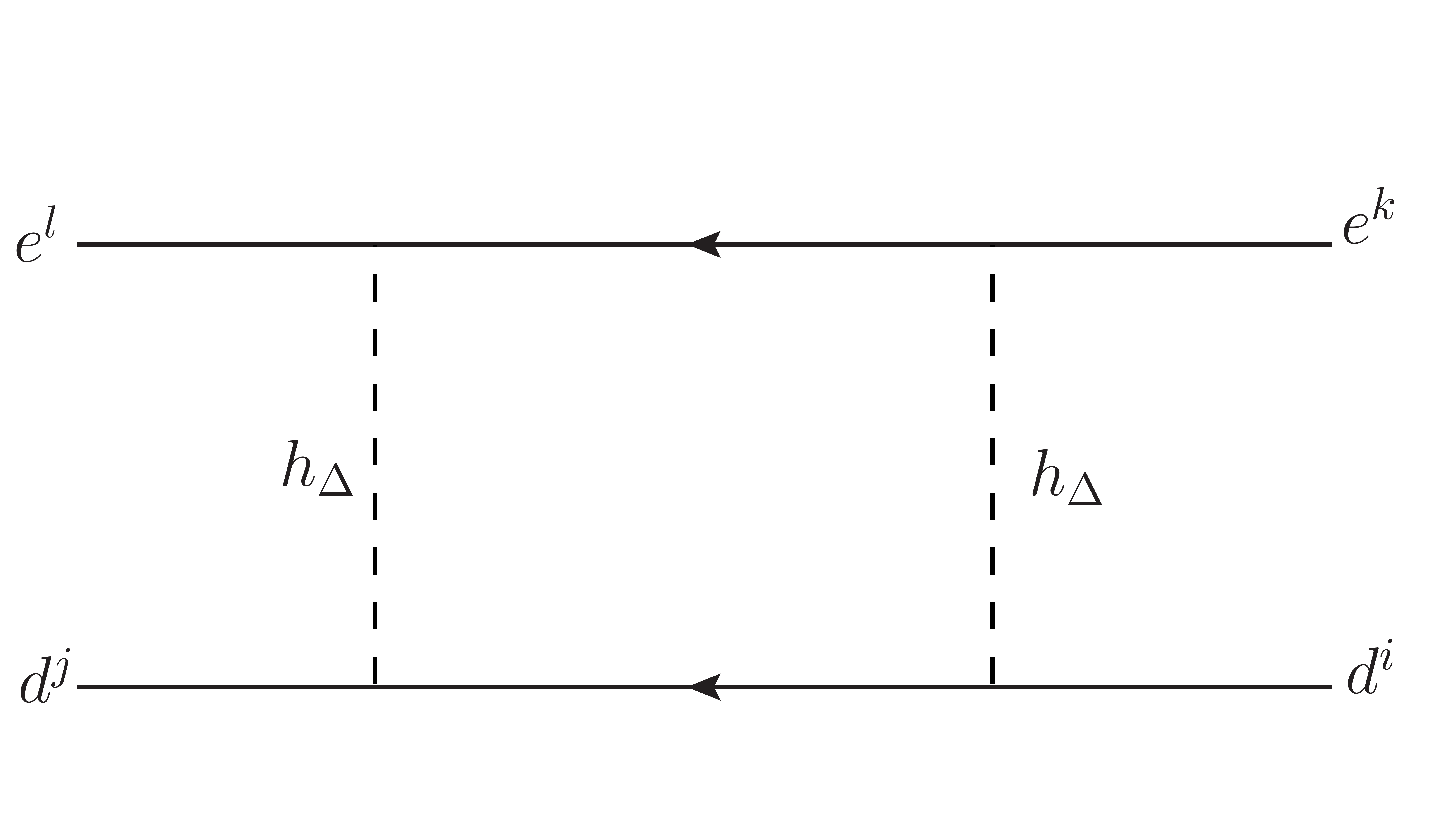}~~~~~~
\includegraphics[scale=0.12]{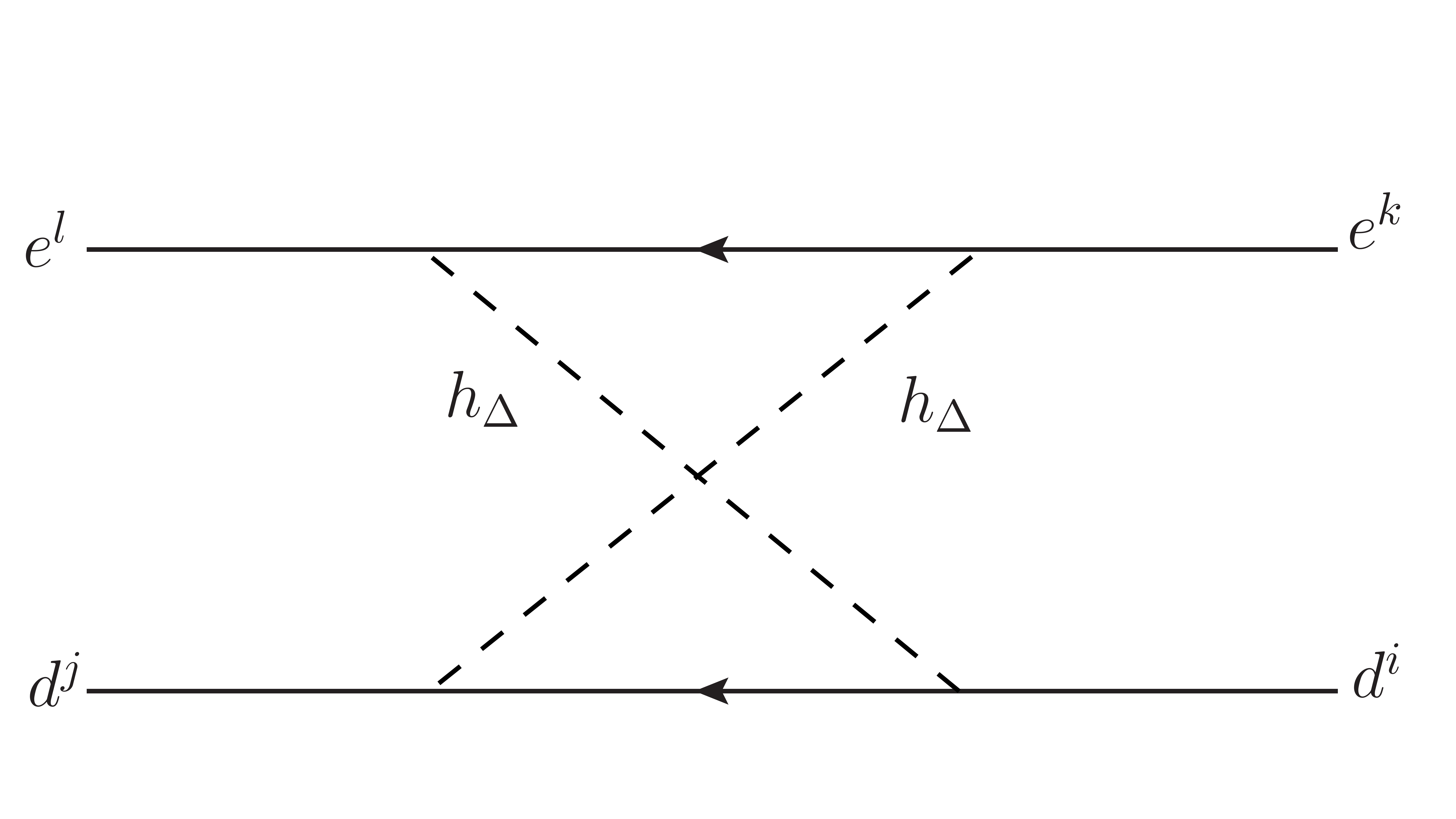}
}
\caption{\label{Box-hh} 
The box diagrams involving only scalars. 
}
\end{figure}

We shall here look at box-diagram contributions to the semi-leptonic operators. 
Let us start with the box diagrams involving two LQs (Fig.\,\ref{Box-XX}). 
Under the condition (i), the LQ couplings to two SM fermions are vanishing (i.e. $X_A=0$), so that we can only consider the vector-like fermions in the loop.
In the diagrams where the internal fermion mass is not picked up, the external fermions maintain the chirality.
Such contribution is given in the form of 
\begin{align}
(g_4)^4 \sum_{I,J \geq4} f (m^e_I,m^d_J;m_X,m_X) 
\bigl[\Omega_A^\dag\bigr]_{l I} \bigl[ \Omega_A \bigr]_{I k} 
\bigl[\Omega_B\bigr]_{j J} \bigl[ \Omega_B^\dag \bigr]_{J i},  
\label{eq:LQbox}
\end{align}
where $A,B=L,R$ and $I,J = 4,5,\cdots,N_L+N_R+3$ run only over the vector-like fermions. 
Given the unitarity of the LQ couplings $\Omega_A$, this equation reminds us of the well-known GIM mechanism. 
Hence, we now suggest another suppression condition,
\begin{itemize}
 \item [(ii)] {vector-like fermion masses are universal:
$m^d_I=: m_D$ and $m^e_I=: m_E$ for $I\ge 4$ .
}  
\end{itemize}
This condition corresponds to $D_{d_L} \simeq m_D \id{N_L}$, $D_{d_R} \simeq m_D \id{N_R}$, 
$D_{e_L} \simeq m_E \id{N_L}$ and $D_{e_R} \simeq m_E \id{N_R}$
up to $\order{\eta}$ contributions.  
Under the conditions (i) and (ii), 
the loop function $f(m^e_I,m^d_J;m_X,m_X) $ can be pulled out of the summation over the internal fermion species 
and then we recognize the GIM-like suppression:
\begin{align}
 \sum_{I\ge 4}  
         \bigl[ \Omega_A^\dag\bigr]_{lI}  \bigl[\Omega_A\bigr]_{Ik} 
\sim \bigl[ V_{L3}^\dag V_{L3} \bigr]_{lk} \sim \delta_{lk} ,
\label{eq:unitary_demo}
\end{align}
where Eq.\,\eqref{eq-unitV3A} is used. 
One can also show that the unitarity assures a similar suppression in the other combination $\sum_{J\ge 4} \bigl[\Omega_B\bigr]_{j J} \bigl[ \Omega_B^\dag \bigr]_{J i} \sim \delta_{ji}$. 
The flavor violating processes with $i\neq j$ or $k\neq l$ are therefore suppressed by the unitarity of $\Omega_A$.

When the internal fermion mass is picked up, the chirality of the external fermions is flipped and different coupling combinations, e.g. $\Omega_L^\dag D_d \Omega_R$, appear in the Wilson coefficients. 
Such contribution contains the factor,
\begin{align}
m_D \sum_{I\ge 4}  
         \bigl[ \Omega_A^\dag\bigr]_{lI}  \bigl[\Omega_{\ol{A}} \bigr]_{Ik} 
\sim 0 ,
\label{eq:unitary_demo2}
\end{align}
where the condition (ii) is assumed.
This result reflects the fact that there is no mixing between $SU(2)_L$ singlet and doublet fermions.
Thus, any sizable flavor violation is not induced via the box diagram involving two LQs once the conditions (i) and (ii) are imposed. 

We next examine the box diagrams with $h_\Delta$ or $\Delta_8$ in the loop.
We have the diagrams involving one LQ and one scalar (Fig.\,\ref{Box-Xh56}) and 
the diagrams involving only scalars (Fig.\,\ref{Box-hh}).
These diagrams generate the semi-leptonic operators, 
\begin{align}
\label{eq-Leffs}
i \Lcal_{\eff}&= 
        C_{VLL}^{\;ij,kl}
        \left( \dbL{j} \gamma_\mu \dL{i}\right) \left( \ebL{l} \gamma_\mu \eL{k}\right) 
       +C_{VRR}^{\;ij,kl}
        \left( \dbR{j} \gamma_\mu \dR{i}\right) \left( \ebR{l} \gamma_\mu \eR{k}\right) 
 \\ 
&\     +C_{VRL}^{\;ij,kl}
        \left( \dbR{j} \gamma_\mu \dR{i}\right) \left( \ebL{l} \gamma_\mu \eL{k}\right) 
       +C_{VLR}^{\;ij,kl}
        \left( \dbL{j} \gamma_\mu \dL{i}\right) \left( \ebR{l} \gamma_\mu \eR{k}\right) 
\notag \\ 
&\ 
       +C_{SLL}^{\;ij,kl} \left(\dbR{j} \dL{i}\right) \left(\ebR{l} \eL{k}\right)
       +C_{SRR}^{\;ij,kl} \left(\dbL{j} \dR{i}\right) \left(\ebL{l} \eR{k}\right)
\notag \\ 
&\     +C_{SRL}^{\;ij,kl} \left(\dbL{j} \dR{i}\right) \left(\ebR{l} \eL{k}\right)
       +C_{SLR}^{\;ij,kl} \left(\dbR{j} \dL{i}\right) \left(\ebL{l} \eR{k}\right) \notag,
\end{align}
where the flavors of the quarks or leptons are different ($i\neq j$ or $k\neq l$). 
Under the conditions (i) and (ii) and assuming $m_{h_\Delta} = m_{\Delta_8}$ for simplicity, the Wilson coefficients are given by 
\begin{align}
C_{VAA}^{\;ij,kl}
=&\ 
\frac{3 m_E^2 m_D^2}{8v_\Delta^4 m_X^2} \Ups_A^{li}\Ups_A^{*\; kj} 
\left[ \frac{3}{16} G_1\left(m_E,m_D; m_X, m_{h_\Delta} \right) 
 - \frac{g_4^2 v_\Delta^2}{m_X^2} G_0\left(m_E,m_D; m_X, m_{h_\Delta} \right) 
 \right],   \\
C_{VA\ol{A}}^{\;ij,kl} =&\ 0, \\ 
C_{SAA}^{\;ij,kl}=&
\frac{9 m_E^3 m_D^3}{512 v_\Delta^4 m_X^4} 
    \Psi_{\ol{A}}^{\;lk} \ol{\Psi}_{\ol{A}}^{\;ji} 
    \tF_0(m_E,m_D;m_{h_\Delta},m_{h_\Delta}),  \\ 
C^{\;ij,kl}_{SA\ol{A}}
=&\ 
\frac{3m_E^2m_D^2}{4 v_\Delta^4 m_X^2} \Ups_{\ol{A}}^{li} \Ups_{A}^{*\; kj}
\left[-\frac{3}{16} G_1(m_E,m_D; m_{X},m_{h_\Delta}) 
+ \frac{g_4^2 v_\Delta^2}{m_X^2} G_0(m_E,m_D; m_{X},m_{h_\Delta}) \right]  \\ \notag
&\ + \frac{m_E^3m_D^3}{8v_\Delta^4m_X^4} \Psi^{\; lk}_{\ol{A}}  \ol{\Psi}^{*\; ji}_{A} 
 \tF_0(m_E,m_D; m_{h_\Delta},m_{h_\Delta}),  
 \label{eq:Semilep_CX}
\end{align}
where the loop functions $\tF_0, G_0$ and $G_1$ are defined in Appendix \ref{app-Int}.
The diagrams with one LQ and one scalar contribute to $C_{VAA}$ and $C_{SA\ol{A}}$, 
while the diagrams with two $h_\Delta$ contribute to $C_{SAA}$ and $C_{SA\ol{A}}$.
Note that the box diagrams involving one LQ and one Goldstone boson are vanishing because of the GIM-like suppression. 

The box contributions 
are expressed in terms of the combinations of the unitary matrices,
\begin{align}
 \Ups_A := \Omega_A^\dag \Ptb\Omega_{\ol{A}} \Ptb \Omega_A^\dag, 
\quad 
 \Psi_A :=  \Omega_A^\dag \Ptb \Omega_{\ol{A}} 
                      \Ptb \Omega^\dag_{A} \Ptb \Omega_{\ol{A}},  
\quad 
 \ol{\Psi}_A:=  \Omega_A\Ptb \Omega^\dag _{\ol{A}} 
                      \Ptb \Omega_{A} \Ptb \Omega^\dag_{\ol{A}},  
\end{align}
where $\Ptb := \mathrm{diag}\left(0,0,0,1,\cdots,1\right)$ 
is a projection matrix to the vector-like families. 
$\Psi$ and $\ol{\Psi}$ originate from the diagrams only with the scalars. 
Using Eq.\,\eqref{eq-OmgLR}, 
the explicit structures are given by
\begin{align}
\label{eq-Psi}
\Psi_L =&\  
\begin{pmatrix}
 0 & 0 & V_{L3}^\dag W_R Y_L^\dag W_R \\ 
 W_L^\dag Y_R W_L^\dag V_{R3} & W_L^\dag Y_R W_L^\dag Y_R & 0 \\ 
 0 & 0 & Y_{L}^\dag W_R Y_L^\dag W_R \\ 
\end{pmatrix},  \\
\Psi_R =&\ 
\begin{pmatrix}
 0 & V_{R3}^\dag W_L Y_R^\dag W_L & 0 \\ 
 0 & Y_R^\dag W_L Y_R^\dag W_L & 0 \\ 
W_R^\dag Y_L W_R^\dag V_{L3} & 0 & W_R^\dag Y_L W_R^\dag Y_L  
\end{pmatrix}, \\ 
\ol{\Psi}_L=&\ 
\begin{pmatrix}
 0 & 0 & V_{3L} W_R^\dag Y_L W_R^\dag \\ 
W_L Y_R^\dag W_L V_{3R}^\dag & W_L Y_R^\dag W_L Y_R^\dag & 0 \\ 
0 & 0 & Y_L W_R^\dag Y_L W_R^\dag 
\end{pmatrix}, \\
\ol{\Psi}_R =&\ 
\begin{pmatrix}
 0 & V_{3R} W_L^\dag Y_R W_L^\dag & 0 \\ 
 0 & Y_{R} W_L^\dag Y_R W_L^\dag & 0 \\ 
W_R Y_L^\dag W_R V_{3L}^\dag  & 0 &W_R Y_L^\dag W_R Y_L^\dag 
\end{pmatrix}. 
\end{align}
Since the top-left $3\times 3$ blocks are zero, 
the flavor violating processes via these coupling structures are suppressed under the conditions (i) and (ii).

The situation is different in the diagrams involving both LQ and adjoint scalars, 
denoted by $\Ups_L$ and $\Ups_R$, which are given by  
\begin{align}
\label{eq-Ups}
 \Ups_L = 
\begin{pmatrix}
 V_{L3}^\dag W_R V_{3L}^\dag & 0 & V_{L3}^\dag W_R Y_L^\dag \\ 
 0 & W_L^\dag Y_R W_L^\dag & 0 \\ 
 Y_{L}^\dag W_R V_{3L}^\dag & 0 & Y_{L}^\dag W_R Y_L^\dag \\ 
\end{pmatrix}, 
\quad 
\Ups_R = 
\begin{pmatrix}
 V_{R3}^\dag W_L V_{3R}^\dag & V_{R3}^\dag W_L Y_R^\dag & 0 \\ 
 Y_R^\dag W_L V_{3R}^\dag & Y_R^\dag W_L Y_R^\dag & 0 \\ 
 0 & 0 & W_R^\dag Y_L W_R^\dag 
\end{pmatrix}.  
\end{align}
We find that the SM top-left blocks are $3\times3$ unitary matrices and cannot be vanishing. 
Note that one of the two indices of $\Ups_{L,R}^{li}$ represents quark flavor and the other represents lepton flavor. 
Thus, even if the SM blocks in $\Ups_{L,R}^{li}$ are diagonal, it does not indicate that the flavor violating processes with $i\neq j$ or $k\neq l$ are vanishing. 
Indeed, we will see in the next section that 
the rapid $K_L \to \mu e$ decay is caused by taking $\Ups_{L,R}^{li} = \delta_{li}$.

The above results can be understood as follows. 
Without the $SU(2)_L$ breaking effects, 
the Yukawa couplings of $\Delta$ with the SM leptons $e_{L,R}$ are schematically given by 
\begin{align}
\Delta \left(\ol{e}_L Y_{e_L}  E_R + \ol{\Ecal}_L Y_{e_R} e_R\right) + h.c.,   
\end{align}
where $E_R$ ($\Ecal_L$) is the $SU(2)_L$ doublet (singlet) heavy lepton. 
Hence, the SM leptons with different chirality cannot participate in 
the same Yukawa couplings of $\Delta$ without the $SU(2)_L$ breaking effects. 
As a result, the non-vanishing box contributions with two scalars are proportional to $Y_{e_L}^\dag Y_{e_L}$ or $Y_{e_R}^\dag Y_{e_R}$ which are diagonal under the condition (ii). 
Thus, those diagrams do not induce the flavor violating interaction. 
On the other hand, 
once the LQ interactions of the SM down quarks $d_{L,R}$ 
\begin{align}
X^\mu \left( g_{d_L} \ol{d}_L \gamma_\mu E_L+g_{d_R} \ol{d}_R \gamma_\mu \Ecal_R\right) 
+ h.c. ,
\end{align}
are considered, the SM down-type quarks can interact with the SM charged leptons via the vector-like leptons. 
Such contribution has the coupling structure of $Y_{e_R} g_{d_R}$ or $Y_{e_L}^\dag g_{d_L}$, which corresponds to $\Ups_{L,R}$, and hence is non-vanishing. 
$\Ups_{L,R}$ is also understood as a generalized version of $g_4 \kappa_L$ in Eq.\,\eqref{eq-dtoe}.

As a side remark, 
we comment on the flavor violation in four-quark and four-lepton operators.
Based on our finding in this section, 
the unsuppressed coupling structures $\Ups_{L,R}$ are obtained only from the box diagrams with one LQ and one $h_\Delta$ (or $\Delta_8$). 
Such diagrams only show up in the $\ol{d_i} d_j\to \ol{e_k} e_l$ processes and do not induce the four-quark and four-lepton operators.
The latter operators are only generated from the box diagrams with two LQs or two $h_\Delta$, 
providing the coupling structures of $\Omega_A \Omega^\dagger_A$, $\Omega_A \Omega^\dagger_{\ol{A}}$, $\Psi$, or $\ol{\Psi}$ which only contain flavor-conserving or vanishing elements for the SM fermions. 
Therefore, the flavor violations from the four-quark and four-lepton operators, 
such as neutral meson mixing 
and three-body lepton flavor violating decays $\ell \to \ell^\prime \ell^\prime \ell^{\prime\prime}$ 
($\ell,\ell^\prime, \ell^{\prime\prime}=e,\mu,\tau$), 
are suppressed under the conditions (i) and (ii).

\section{Phenomenology}
\label{sec-pheno}

We evaluate the LFV processes, especially the leptonic meson decays and $\mu\to e$ conversion, assuming the condition (i) and (ii).
We see that there is an upper limit on the vector-like fermion mass for a given LQ mass when experimental bounds on those processes are respected.

\subsection{Flavor violating leptonic decays of neutral mesons}  
\begin{table}[t]
 \centering
\caption{\label{tab-valmsn} 
Values of masses $m_M$, lifetimes $\tau_M$ and decay constants $f_M$ 
of the mesons $M=K,B_d,B_s$~\cite{Zyla:2020zbs,FlavourLatticeAveragingGroup:2019iem}.
}
\begin{tabular}[t]{c|ccc} \hline 
  $M$  &  $m_M$ [$\GeV$] & $\tau_M$ [$\times 10^{12}\cdot \GeV^{-1}$] & $f_M$ [$\GeV$] \\ \hline \hline 
 $K$     & 0.4976 & $7.773\times 10^{4}$ & 0.1552 \\
 $B_d$   & 5.280  & $2.308$               & 0.1920 \\   
 $B_s$   & 5.367  & $2.320$               &0.2284  \\   
\hline 
\end{tabular}
\end{table}

Rare meson decays $M_{ij} \to e_k^- e_l^+$, where $M_{ij}$ is a meson composed of $\ol{d}_j d_i$, are the key processes induced from the semi-leptonic operators of our interest. 
In general, the partial decay width of $M_{ij} \to e_k^- e_l^+$ is given by
\begin{align}
 \Gamma(M_{ij} \to e_k^- e_l^+) 
 = \frac{f_M^2 \beta}{16\pi m_M} 
   \left[
    \left( m_M^2-m_l^2 - m_k^2 \right) a^{ij,kl} - b^{ij,kl} 
   \right],  
\end{align}
\begin{align}
 a^{ij,kl} =&\ 
\sum_{A=L,R} \abs{\ol{m}_M S_A^{ij,kl} - m_l V_A^{ij,kl} + m_k V_{\ol{A}}^{ij,kl}}^2,  \\ 
 b^{ij,kl}=&\ 4 m_k m_l \mathrm{Re}\left[ 
 \left(\ol{m}_M S_L^{ij,kl} - m_l V_L^{ij,kl} + m_k V_R^{ij,kl}\right) 
 \left( \ol{m}_M S_R^{ij,kl} - m_l V_R^{ij,kl} + m_k V_L^{ij,kl} \right)^* 
  \right],
\end{align}
\begin{align}
V_A^{ij,kl} := \frac{C_{VRA}^{ij,kl}- C_{VLA}^{ij,kl}}{2}, \quad 
S_A^{ij,kl} := \frac{C_{SRA}^{ij,kl}- C_{SLA}^{ij,kl}}{2}, 
\end{align}
where $f_M$ and $m_M$ are a decay constant and mass of a meson $M$, 
and  
\begin{align}
 \beta := \sqrt{1-2\frac{m_l^2+m_k^2}{m_M^2} + \frac{(m_l^2-m_k^2)^2}{m_M^4}}. 
\end{align}
Here, we used 
\begin{align}
 \bra{0} \ol{d}_j \gamma^\mu \gamma_5 d_i \ket{M_{ij}} = i f_M P_M^\mu, 
\quad   
 \bra{0} \ol{d}_j \gamma_5 d_i \ket{M_{ij}} = - i f_M \ol{m}_M,   
\label{eq-fMij}
\end{align}
where $P^\mu_M$ is a four momentum of a meson $M$ 
and $\ol{m}_M := {m_M^2}/({m_{d_i} + m_{d_j}})$ with $m_{d_i}$ 
being mass of quark $d_i$.  
\begin{table}[t]
 \centering
\caption{\label{tab-DecayModes}
LFV decay modes of neutral mesons. 
The leptonic indices $(k,l)$ are added with its counterpart $(l,k)$. 
}
\begin{tabular}[t]{c|ccc} \hline
observable                   &   upper limit& $(i,j),~(k,l)$ &Ref. \\ \hline \hline
$\br{K_L}{\mu e}$     & $4.7\times 10^{-12}$& $(1,2),~(1,2)$ & \cite{Zyla:2020zbs}\\ 
\hline 
$\br{B_d}{\mu e}$     & $1.0\times 10^{-9}$& $(1,3),~(1,2)$ & \cite{Zyla:2020zbs}\\
$\br{B_d}{\tau e}$    & $2.8\times 10^{-5}$& $(1,3),~(1,3)$ & \cite{Zyla:2020zbs}\\
$\br{B_d}{\tau \mu}$  & $1.4\times 10^{-5}$& $(1,3),~(2,3)$ & \cite{Zyla:2020zbs}\\
\hline
$\br{B_s}{\mu e}$     & $5.4\times 10^{-9}$& $(2,3),~(1,2)$ & \cite{Zyla:2020zbs}\\
$\br{B_s}{\tau e}$    & -& $(2,3),~(1,3)$ & \cite{Zyla:2020zbs}\\
$\br{B_s}{\tau \mu}$  & $4.2\times 10^{-5}$& $(2,3),~(2,3)$ & \cite{Zyla:2020zbs}\\
\hline 
\end{tabular}
\end{table}
Since the LQ is much heavy compared to the meson, 
the RG corrections of the strong coupling constant are included 
by replacing in Eq.\,\eqref{eq-fMij},~\cite{Dolan:2020doe} 
\begin{align}
 \ol{m}_M \to \ol{m}_M  R_M(m_X),  
\end{align} 
where 
\begin{align}
 R_M(m_X) := R\left({m_M, m_c; 3}\right) 
             R\left({m_c, m_b; 4}\right) 
             R\left({m_b, m_t; 5}\right)
             R\left({m_t, m_X; 6}\right),   
\end{align}
with 
\begin{align}
 R(\mu_1, \mu_2; n_f) := \left(\frac{g_3(\mu_1)}{g_3(\mu_2)}\right)^{\frac{8}{11-2n_f/3}}.
\end{align}
Here we assume that all the new particles are much heavier than the top quark and 
as heavy as $m_X$.
The branching fractions are given by 
\begin{align}
 \br{M_{ij}}{e_k e_l} \simeq \tau_M 
  \Bigl\{\Gamma(M_{ij}\to e_k^- e_l^+) + \Gamma\left(M_{ij}\to e_k^+ e_l^-\right) \Bigr\}. 
\end{align} 
In our calculation, we use the values of the meson parameters and the experimental upper bounds in Tables \ref{tab-valmsn} and \ref{tab-DecayModes}, respectively.

\subsection{$\mu$-$e$ conversion}  

When the quark flavor diagonal pieces in Eq.\,(\ref{eq-Leffs}) are non-vanishing, 
$\mu$-$e$ conversion can also provide a leading constraint.
The conversion rate is given by~\cite{Cirigliano:2009bz}
\begin{align}
\Gamma_{\mathrm{conv}} = 4 m_\mu^5
               \left(
                   \abs{  \sum_{N=p,n} \left(\tilde{C}_{VL}^N V_N+   \left. m_N \tilde{C}_{SL}^N S_N  \right. \right)  }^2  +
                   \abs{ \sum_{N=p,n} \left( \tilde{C}_{VR}^N V_N +   \left. m_N \tilde{C}_{SR}^N S_N \right.\right)   }^2
               \right),
\end{align}
where
\begin{align}
\tilde{C}_{VL}^N = \sum_{q=u,d,s} C^q_{VL} f_{V_N}^q,\quad
\tilde{C}_{SL}^N = \sum_{q=u,d,s} C^q_{SL} f_{S_N}^q + \frac{2}{27} f_G^N \sum_{Q=c,b,t} C^Q_{SL}, \\
\tilde{C}_{VR}^N = \sum_{q=u,d,s} C^q_{VR} f_{V_N}^q,\quad
\tilde{C}_{SR}^N = \sum_{q=u,d,s} C^q_{SR} f_{S_N}^q + \frac{2}{27} f_{G}^N \sum_{Q=c,b,t} C^Q_{SR}.
\end{align}
The values of nucleon form factors for light quarks are collected in Table~\ref{tab-constmue}. 
The form factor for gluon is related to those for light quarks via the QCD trace anomaly, $f_{G}^N = 1- \sum_{q=u,d,s} f_{S_N}^q$.
We ignore the vector-like quark contributions to $\tilde{C}^{N}_{SL}$ and $\tilde{C}^{N}_{SR}$ 
through the trace anomaly since these are suppressed by the vector-like quark masses.
In our model, the scalar and vector coefficients are loop-induced and given in terms of the Wilson coefficients of the semi-leptonic operators given in Eq.(\ref{eq-Leffs}), 
\begin{align}
 C^{d_i}_{VL} &= \frac{1}{2} (C_{VLL}^{ii,21}+C_{VRL}^{ii,21}), \quad 
 C^{d_i}_{VR} = \frac{1}{2} (C_{VRR}^{ii,21}+C_{VLR}^{ii,21}), \\
 C^{d_i}_{SL} &= \frac{1}{2m_{d_i}} (C_{SLR}^{ii,21}+C_{SRR}^{ii,21}), \quad
 C^{d_i}_{SR} =\frac{1}{2m_{d_i}} (C_{SRL}^{ii,21}+C_{SLL}^{ii,21}),
\end{align}
where the index $i$ is not summed. 
The current (future) limit is set on the conversion rate per capture rate~\cite{Bertl:2006up,Natori:2014yba,Kuno:2013mha,Abrams:2012er}, 
\begin{align}
 \br{\mu}{e}^{\mathrm{Au(Al)}} = \frac{\Gamma_{\mathrm{conv}}}{\Gamma_{\mathrm{capt}}}
       < 7\times 10^{-13}~\left(6\times 10^{-17}\right).
\end{align}

\begin{table}[t]
\centering
\caption{\label{tab-constmue}
Values of vector~\cite{Cirigliano:2009bz}  and scalar~\cite{Junnarkar:2013ac} nucleon form factors.
The coefficients $S_N$, $V_N$ where $N=p,\,n$ are calculated in Ref.~\cite{Kitano:2002mt}.
The capture rates are given in Ref.~\cite{Kitano:2002mt,Suzuki:1987jf}.
}
\begin{tabular}[t]{ccccc} \hline
 $f_{V_p}^u$&  $f_{V_p}^d$ &  $f_{V_n}^u$&  $f_{V_n}^d$ &  $f_{V_p}^s = f_{V_n}^s$ \\ \hline
$2$ & $1$ & $1$ & $2$ & $0$ \\  \hline \hline
 $f_{S_p}^u$&  $f_{S_p}^d$ &  $f_{S_n}^u$&  $f_{S_n}^d$ &  $f_{S_p}^s = f_{S_n}^s$ \\ \hline
$0.0191$ & $0.0363$ & $0.0171$ & $0.0404$ & $0.043$ \\
\hline
& \\
\end{tabular}
\begin{tabular}[t]{c|cccc|c} \hline
Target & $S_p$ & $S_n$ & $V_p$ & $V_n$ & $\Gamma_{\mathrm{capt}}~[10^{6}\cdot s^{-1}]$\\ \hline\hline
Au      & $0.0614$ & $0.0918$ & $0.0974$ & $0.146$   & $13.07$ \\
Al       & $0.0155$ & $0.0167$ & $0.0161$ & $0.0173$ & $0.705$ \\
\hline
\end{tabular}
\end{table}

\subsection{Simplified analysis} 

We shall compare the box-induced LFV processes with the experimental limits.
For concreteness, we consider $N_L = N_R= 3$ 
which is the minimal option to realize $X_L = X_R = 0$.  
We neglect the sub-dominant effects suppressed by $\eta$, 
and thus all flavor violating processes are induced via $\Ups_{L,R}^{ij}$ which corresponds to the SM blocks of $\Ups_{L,R}$.
In this case,
$\Ups_{L,R}^{ij}$ are $3\times 3$ unitary matrices and are treated as free parameters in our study. 
We further assume that the SM down-type fermions are in the mass basis 
for a given $\Ups_{L,R}^{ij}$, i.e. 
\begin{align}
 m_{33} \simeq \mathrm{diag}\left( m_d, m_s, m_b \right), 
\quad 
\tm_{33} \simeq V_{L3}^\dag m_{LR} V_{R3} 
         \simeq \mathrm{diag}\left( m_e, m_\mu, m_\tau \right).  
\end{align}
We consider for simplicity the relations between the mass parameters 
\begin{align}
m_\VL := m_E = m_D, \quad 
m_{h_\Delta} = m_X ,
\end{align}
where the LQ mass $m_X$ is related to $v_\Delta$ and $g_4$ via Eq.\,\eqref{eq;LQmass}. 
In our analysis, the input parameters are thus
\begin{align}
m_X,\quad  m_\VL,\quad \Ups_L^{ij},\quad \Ups_R^{ij}. 
\end{align}
and $g_4=1$ is fixed, which is consistent with the strong coupling constant at the TeV scale.

Among the LFV meson decays, $K_L\to \mu e$ is the most sensitive to new physics contributions. 
The branching fraction is given by 
\begin{align}
 \br{K_L}{\mu e} 
\simeq \frac{\tau_K m_K  f_K^2}{16\pi m_X^4} \abs{C_0}^2  \left(1-\frac{m_\mu^2}{m_K^2}\right)^2 
\sum_{A=L,R} \sum_{p=1,2} \abs{\Ups^{2p}_A}^2  
                 \abs{\ol{m}_K\Ups_{\ol{A}}^{1\ol{p}}-\frac{m_\mu}{2}\Ups_A^{1\ol{p}}}^2,
\end{align}
where $\ol{p} = 2,1$ for $p=1,2$ and the electron mass is neglected.  
The coefficient $C_0$ is defined as  
\begin{align}
 C_0 := \frac{9m_\VL^4}{16v_\Delta^4} 
         \left(\frac{1}{4}G_1(m_\VL, m_\VL; m_X, m_X)
               - G_0(m_\VL, m_\VL; m_X, m_X ) \right). 
\end{align}
With the sizable chiral enhancement proportional to $\ol{m}_K$, we find the branching fraction to be
\begin{align}
 \br{K_L}{\mu e} \simeq 3.5\times 10^{-5}  \times  \abs{\Ups^{2p}_{A} \Ups^{1\ol{p}}_{\ol{A}}}^2 
                    \left(\frac{\abs{C_0}}{1/(16\pi^2)}  \right)^2 
                    \left(\frac{5~\TeV}{m_X}\right)^4,  
\end{align} 
while without the chiral enhancement, 
\begin{align}
 \br{K_L}{\mu e} \simeq 3.0\times10^{-9} \times 
               \abs{\Ups^{2p}_{A} \Ups^{1\ol{p}}_{A}}^2 
               \left(\frac{\abs{C_0}}{1/(16\pi^2)}  \right)^2 
               \left(\frac{5~\TeV}{m_X}\right)^4.  
\end{align} 
Hence there should be $\order{10^{-7}}$ and $\order{10^{-3}}$ suppression 
from the couplings and masses with and without the chiral enhancement, respectively. 
It is illuminating to give the scaling of the loop function $C_0$ with $m_{\VL} \ll m_X$, 
\begin{align}
|C_0|\simeq \frac{m_\VL^4}{8\pi^2 m_X^4}\left(\frac{7}{4}+{\rm{log}}\frac{m_\VL^2}{m_X^2}\right).
\end{align}
In this limit, the branching fraction is proportional to $(m_\VL/m_X)^8$. 
The mass splitting of the LQ and the vector-like fermions will help to suppress the branching fraction.

The $\mu$-$e$ conversion rate per capture rate on the gold target is given by 
\beq
\br{\mu}{e}^{\mathrm{Au}} = \frac{m_\mu^5}{\Gamma_{\mathrm{capt}}^{\mathrm{Au}} m_X^4} |C_0|^2
\left( \sum_{A=L,R} |\Ups_{A}^{2i}|^2 \left| \sum_{N=p,n} \sum_{i=1,2,3} \frac{1}{2} \Ups_{A}^{1i} f^{d_i}_{V_N} V_N - \frac{m_N}{m_{d_i}} \Ups_{\ol{A}}^{1i} f^{d_i}_{S_N} S_N \right|^2 \right) ,
\eeq
where $f_{V_N}^b=0$ and $f_{S_N}^b=2 f_G^N/27$. 
In the vector-dominant case, i.e. $|\Ups_{A}^{2i} \Ups_{\ol{A}}^{1i}| \ll |\Ups_{A}^{21} \Ups_{A}^{11}|$, 
we have
\begin{align}
\br{\mu}{e}^{\mathrm{Au}} \simeq 3.7 \times 10^{-9} \times \left(\frac{|C_0|}{1/(16\pi^2)}\right)^2
\left(\frac{5\,\TeV}{m_X}\right)^4 \sum_{A=L,R} |\Ups_{A}^{21} \Ups_{A}^{11}|^2 ,
\end{align}
and in the scalar-dominant case, i.e. $|\Ups_{A}^{2i} \Ups_{\ol{A}}^{1i}| \gg |\Ups_{A}^{21} \Ups_{A}^{11}|$, 
\begin{align}
\br{\mu}{e}^{\mathrm{Au}} \sim 7 \times 10^{-7} \times \left(\frac{|C_0|}{1/(16\pi^2)}\right)^2
\left(\frac{5\,\TeV}{m_X}\right)^4 \sum_{A=L,R} \left| \sum_i \Ups_{A}^{2i} \Ups_{\ol{A}}^{1i} \, \frac{f_{S_p}^{d_i}/m_{d_i}}{f_{S_p}^d/m_d} \right|^2 ,
\end{align}
where $f_{S_p}^{d_i} \sim f_{S_n}^{d_i}$ is used.
It should be noted that the lighter quark has larger contribution in the scalar-dominant case 
due in part to a factor of $1/m_{d_i}$.

\begin{figure}[t]
\centering
\begin{minipage}[c]{0.48\hsize}
\centering
\includegraphics[height=80mm]{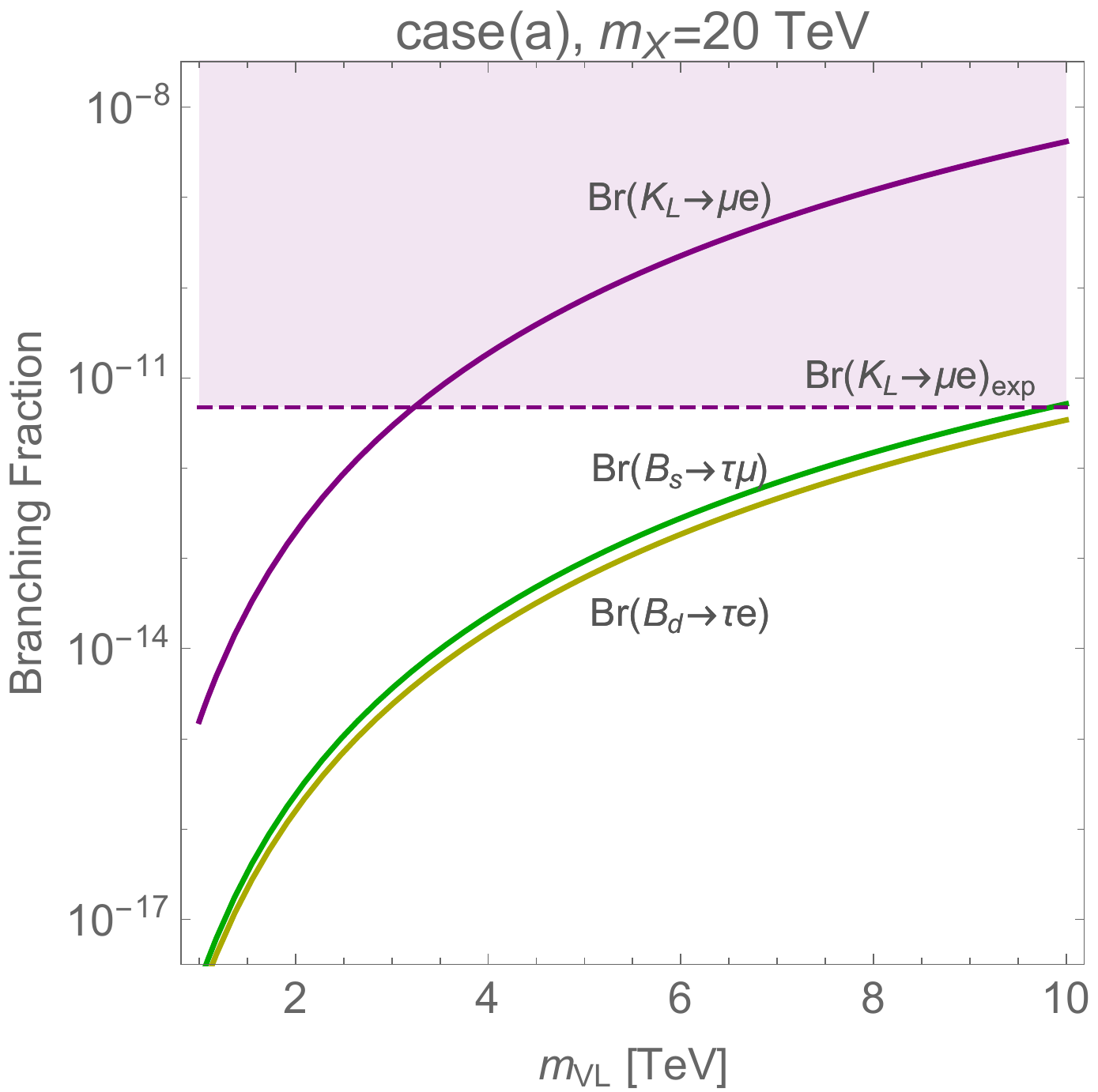}
\end{minipage}
\begin{minipage}[c]{0.48\hsize}
\centering
\includegraphics[height=80mm]{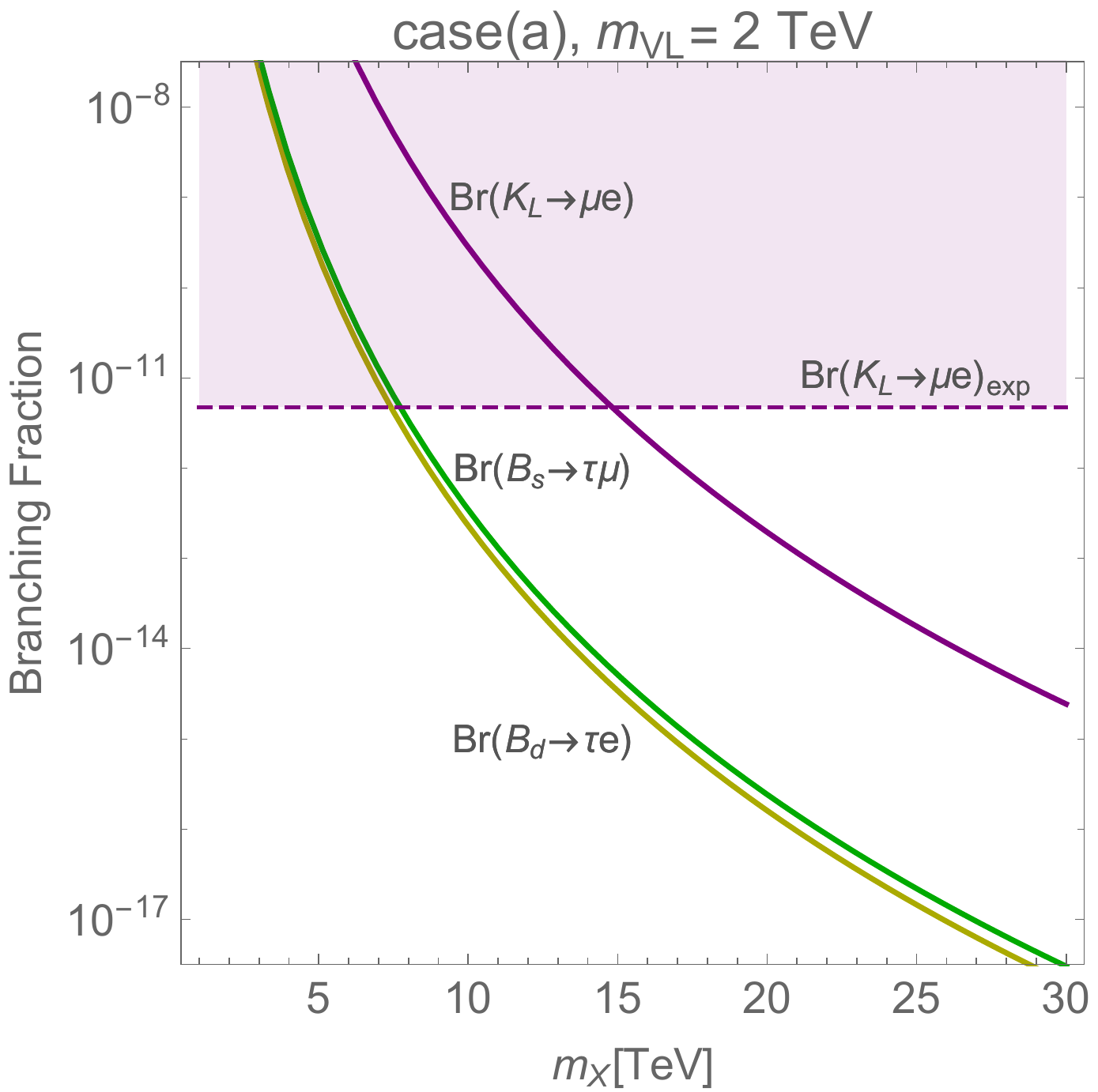} 
\end{minipage}
\caption{\label{fig-simpA}
Values of the branching fractions in the case (a). See the main text for the detail.
}
\end{figure}
 
 \begin{figure}[t]
\centering
\begin{minipage}[c]{0.48\hsize}
\centering
\includegraphics[height=80mm]{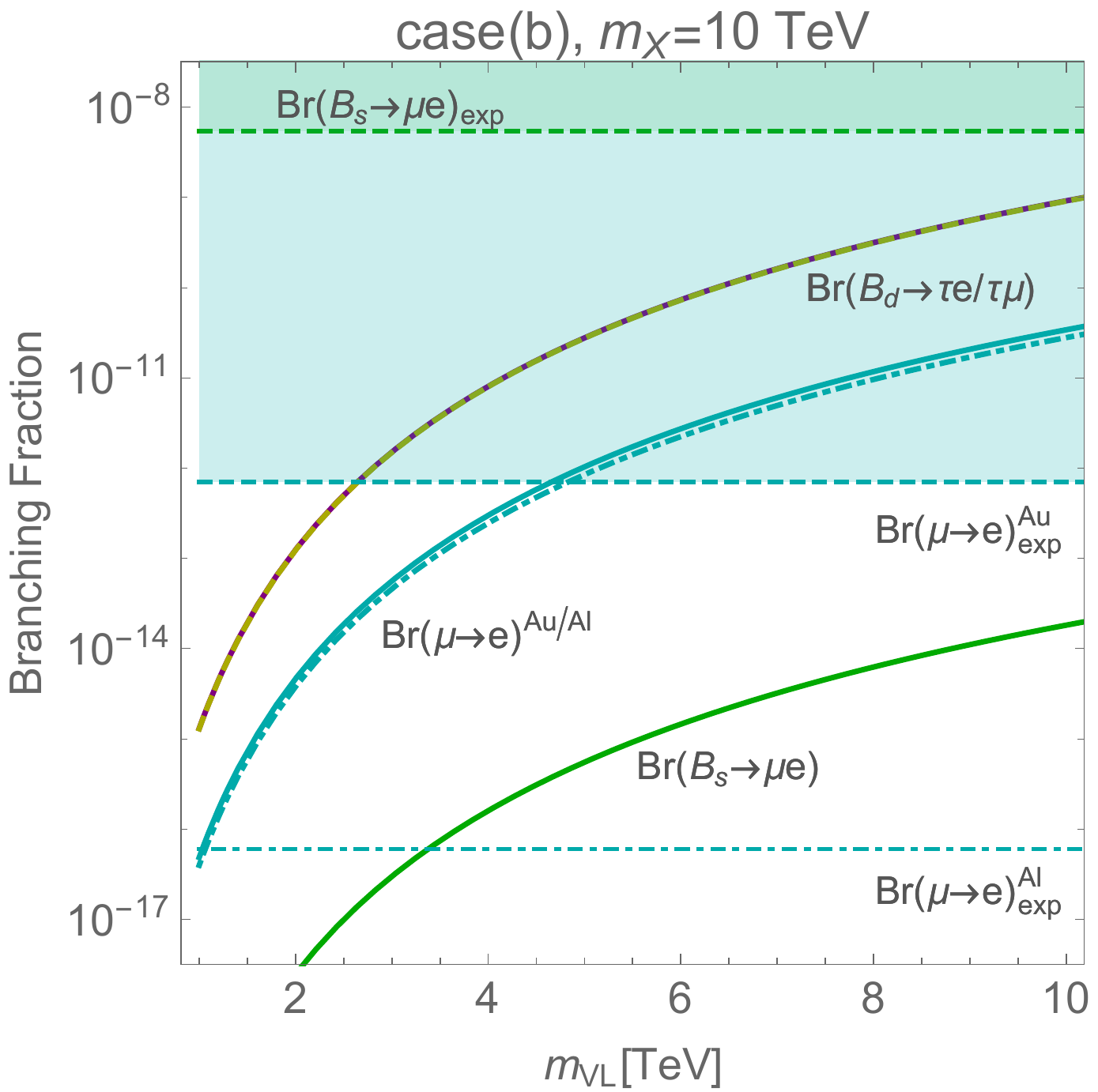}
\end{minipage}
\begin{minipage}[c]{0.48\hsize}
\centering
\includegraphics[height=80mm]{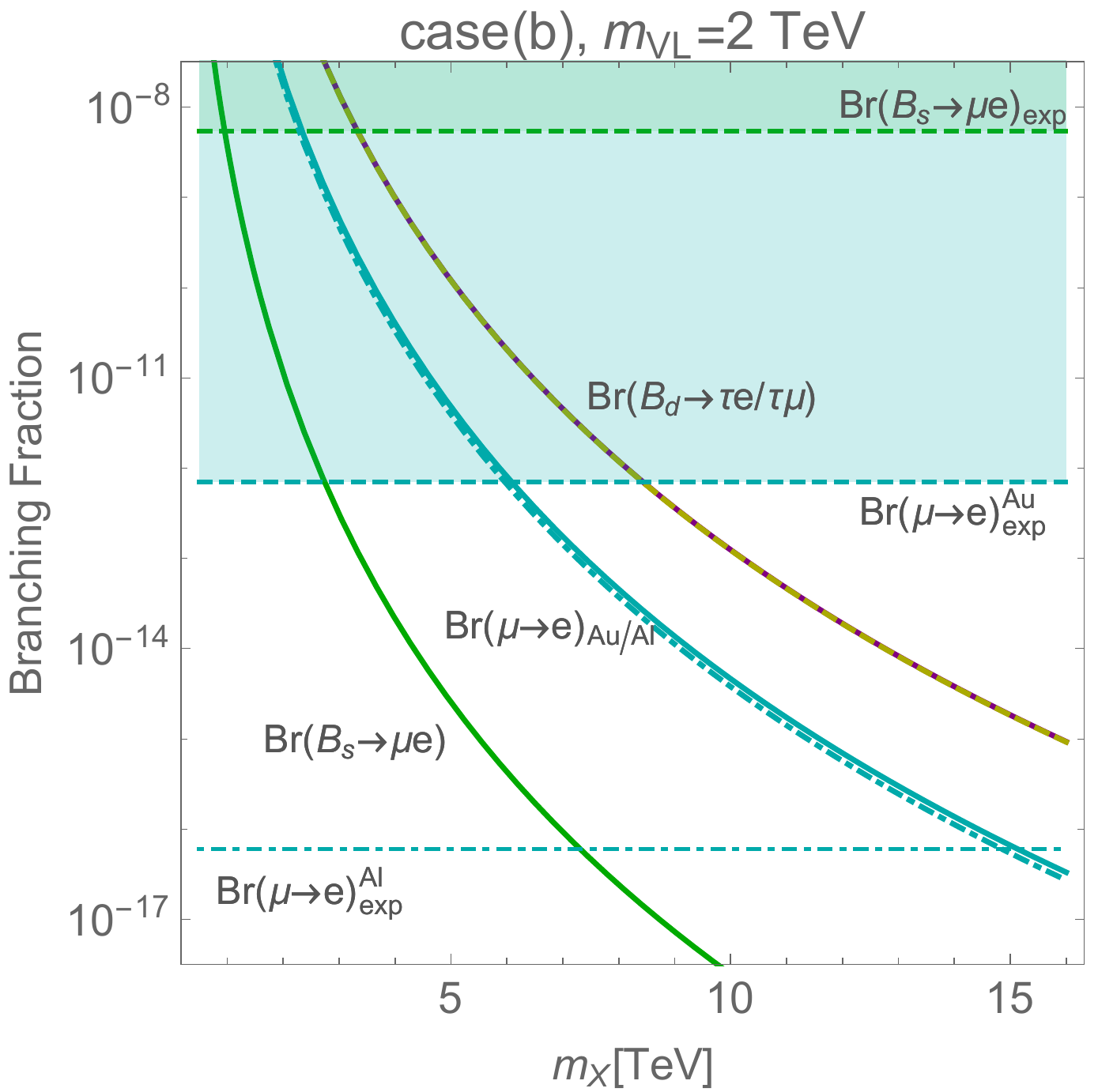} 
\end{minipage}
\caption{\label{fig-simpB}
Values of the branching fractions in the case (b). 
$\br{B_d}{\tau e}\simeq \br{B_d}{\tau\mu}$ and $\br{\mu}{e}^{\mathrm{Au}} \simeq \br{\mu}{e}^\mathrm{Al}$ 
in this case. 
}
\end{figure}

 \begin{figure}[t]
\centering
\begin{minipage}[c]{0.48\hsize}
\centering
\includegraphics[height=80mm]{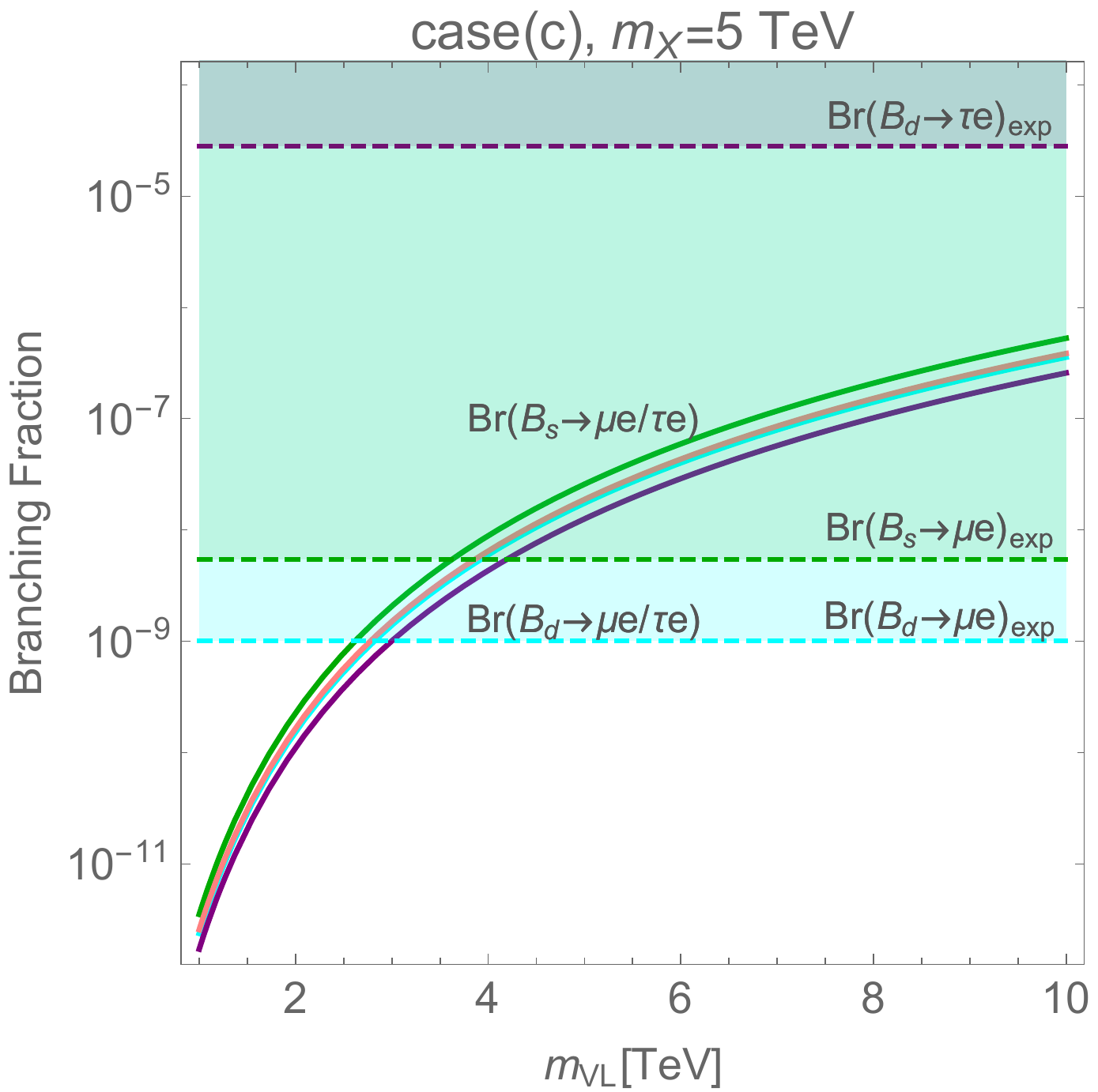}
\end{minipage}
\begin{minipage}[c]{0.48\hsize}
\centering
\includegraphics[height=80mm]{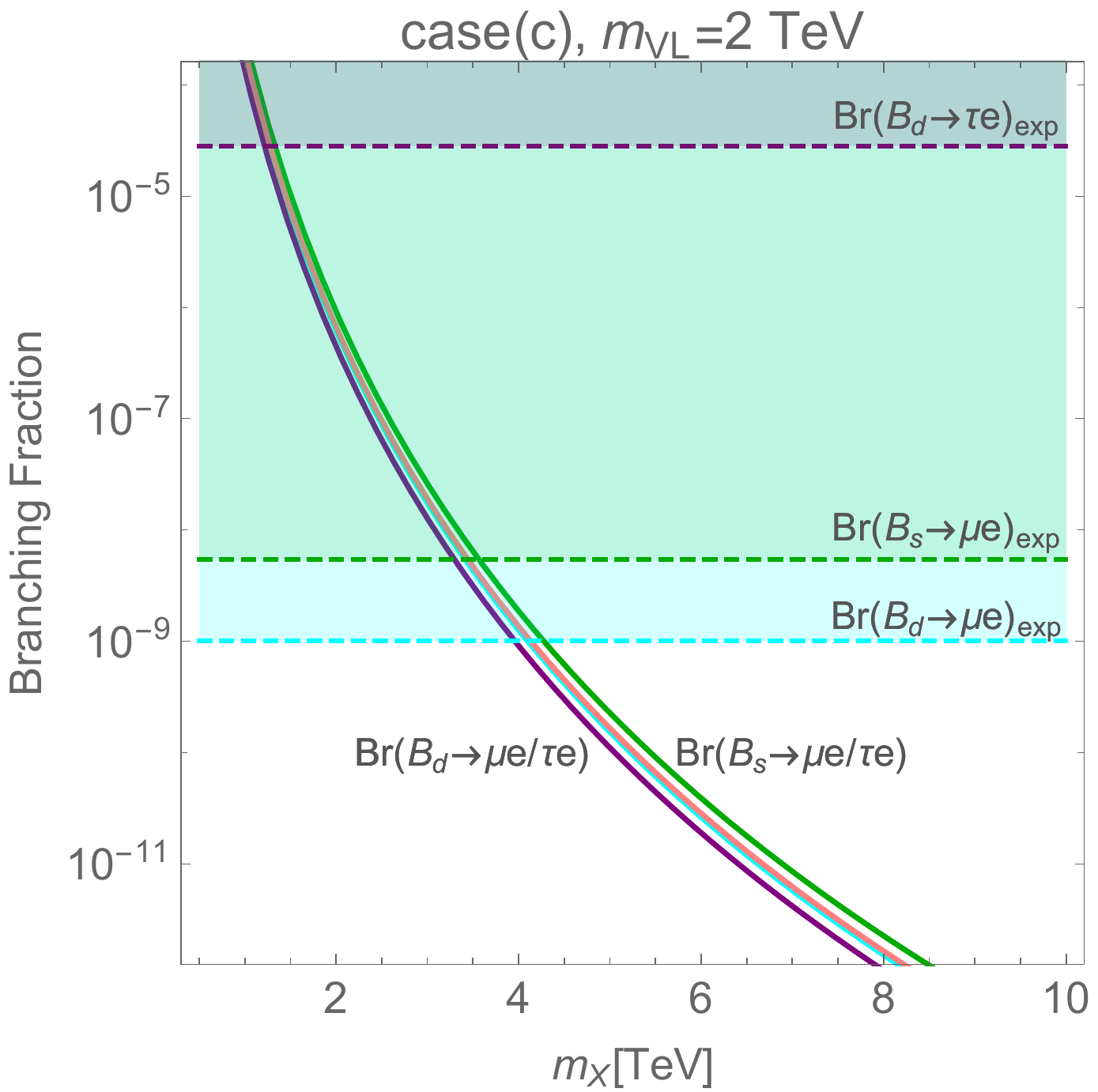} 
\end{minipage}
\caption{\label{fig-simpC}
Values of the branching fractions in the case (c).
}
\end{figure}

We consider three simplified cases: 
\begin{align}
\label{eq-caseAB}
\mathrm{(a)}~~\Ups_L^{ij} = \Ups_R^{ij} = \delta_{ij}, 
\quad\quad 
\mathrm{(b)}~~\Ups_L^{ij} =
\begin{pmatrix}
 0 & 1 & 0 \\ 
 0 & 0 & 1 \\ 
 1 & 0 & 0 \\ 
\end{pmatrix}, 
~\Ups_R^{ij} =  
\begin{pmatrix}
 0 & 0 & 1 \\ 
 0 & 1 & 0 \\ 
 1 & 0 & 0 \\ 
\end{pmatrix},
\end{align}
and
\begin{align}
\mathrm{(c)}~~\Ups_L^{ij} =
\begin{pmatrix}
 0 & 0 & 1 \\ 
 1 & 0 & 0 \\ 
 0 & 1 & 0 \\ 
\end{pmatrix}, 
~\Ups_R^{ij} =  
\begin{pmatrix}
 0 & 0 & 1 \\ 
 0 & 1 & 0 \\ 
 1 & 0 & 0 \\ 
\end{pmatrix}. 
\end{align}
In the case (a), 
there is a large contribution to $K_L \to \mu e$ via $\Ups^{11}_A\Ups^{22}_{\ol{A}}$, 
while $\mu$-$e$ conversion is not induced because $\Ups_{A}^{2i} \Ups_{A}^{1i}=\Ups_{A}^{2i} \Ups_{\ol{A}}^{1i}=0$. 
In the case (b), $K_L\to \mu e$ and $B_d \to \mu e$ are not induced 
and, moreover, the chiral enhanced contributions to $B_{s}\to \mu e$ are vanishing, 
which suggests that the LFV meson decays only provide weak constraints.
By contrast, the $\mu$-$e$ conversion process is induced in this case since $\Ups_L^{23} \Ups_R^{13} = \Ups_R^{22} \Ups_L^{12} =1$.
In the case (c), $K_L\to \mu e$ and $\mu\to e$ conversion are absent 
while $B_d,B_s\to \mu e$ are chiral enhanced.
Since $K_L \to \mu e$ and $\mu\to e$ conversion give much stronger bounds than the others, 
the limits on the case (c) will be the weakest. 

Figure~\ref{fig-simpA} shows the values of the branching fractions in the case (a) 
as a function of $m_X$ ($m_\VL$) in the left (right) panel with $m_\VL = 20~\TeV$ ($m_X = 2~\TeV$). 
The solid lines are our predictions in this model, 
and the horizontal dashed lines are the experimental upper limits. 
Note that the other decay modes not shown in the figures 
are vanishing in this analysis.
It follows from Fig.~\ref{fig-simpA} (left) that,  
with $m_X = 20~\TeV$, we find an upper bound on $m_\VL \lesssim 3~\TeV$ from $\br{K_L}{\mu e}$, 
whereas $\br{B_d}{\tau\mu}$ and $\br{B_s}{\tau e}$ are much smaller 
than the experimental limits of $\order{10^{-5}}$. 
It is remarkable that the branching fractions are suppressed by a factor of $m_\VL^4/v_\Delta^4$ in $C_0$, 
providing an $upper$ bound on the vector-like mass for a given LQ mass. 
When we instead fix the vector-like fermion mass, the LQ mass scale is limit from below. 
One can see in Fig.~\ref{fig-simpA} (right) that the LQ mass should be heavier than $15~\TeV$ with $m_\VL = 2~\TeV$.

Figures~\ref{fig-simpB} and~\ref{fig-simpC} are similar plots to Fig.~\ref{fig-simpA} 
for the case (b) and (c), respectively. 
In the case (b), 
the values of $\br{B_d}{\tau e}$ (purple) and $\br{B_d}{\tau\mu}$ (yellow dashed) 
are degenerate since the muon mass is still negligible compared to $m_b$ and $m_\tau$. 
As regards the $\mu\to e$ conversion, $\br{\mu}{e}^{\mathrm{Au}}$ (cyan) is slightly larger than $\br{\mu}{e}^\mathrm{Al}$ (cyan dashed). 
The horizontal dashed lines are the current upper limits on the corresponding LFV processes
and the horizontal dot-dashed line is the future sensitivity to $\br{\mu}{e}^\mathrm{Al}$. 
The $\mu\to e$ conversion provides the strongest constraint in the case (b), 
and it requires $m_\VL \lesssim 4.5~\TeV$ for $m_X=10~\TeV$ while $m_X \gtrsim 6~\TeV$ for $m_\VL = 2~\TeV$. 
The future $\br{\mu}{e}^\mathrm{Al}$ measurement will improve the limits to  
$m_\VL \lesssim 1~\TeV$ and $m_X \gtrsim 15~\TeV$, respectively. 
We see the complementarity of LFV meson decays and the $\mu\to e$ conversion in the case (a) and (b), however, there is no contribution to both $K_L \to \mu e$ and $\mu \to e$ conversion in the case (c).
The green, pink, cyan and purple lines 
are $\br{B_s}{\mu e}$, $\br{B_s}{\tau_e}$, 
$\br{B_d}{\mu e}$ and $\br{B_d}{\tau e}$, respectively.
The current limit reads $m_\VL \lesssim 3~\TeV$ for $m_X = 5~\TeV$ while $m_X \gtrsim 4~\TeV$ for $m_\VL = 2~\TeV$. 
Thus $5~\TeV$ LQ is not excluded in this case when the VL fermions are sufficiently light. 
This leaves the possibility of resolving the $B$ anomalies~\cite{Iguro:2021kdw}.

Let us comment on the LHC constraints on the model. 
In our analysis, we consider the vector-like fermions to be heavier than 2 TeV 
which is significantly higher than the current LHC limits $m_\VL \gtrsim 1\,\TeV$~\cite{ATLAS:2018cjd,CMS:2019afi}.  
Thus we need higher energy colliders, e.g. $\sqrt{s} = 100~\TeV$ 
to explore the allowed parameter space 
\cite{Altmannshofer:2013zba,Bhattiprolu:2019vdu,Guedes:2021oqx}. 
Besides, there is a model-dependent bound on the light LQ scenario, 
given the existence of a new neutral gauge boson $Z^\prime$ whose mass is correlated to the LQ mass to some extent. 
In fact, $2~\TeV$ LQ is excluded by the di-muon resonance search for $Z^\prime$~\cite{ATLAS:2019erb}, 
when the residual $SU(2)_R \times U(1)_{B-L}$ symmetry is broken by a scalar field with $(\ol{\bf 10}, {\bf 1}, {\bf 3})$ under the PS symmetry~\cite{Iguro:2021kdw}. 
Thus, additional careful consideration will be needed to achieve the LQ lighter than a few TeV. 

\subsection{Comments on penguin diagrams} 
\label{sec-penguin}

\begin{figure}[t]
\centerline{\includegraphics[viewport=70 300 530 770, clip=true, scale=0.7]{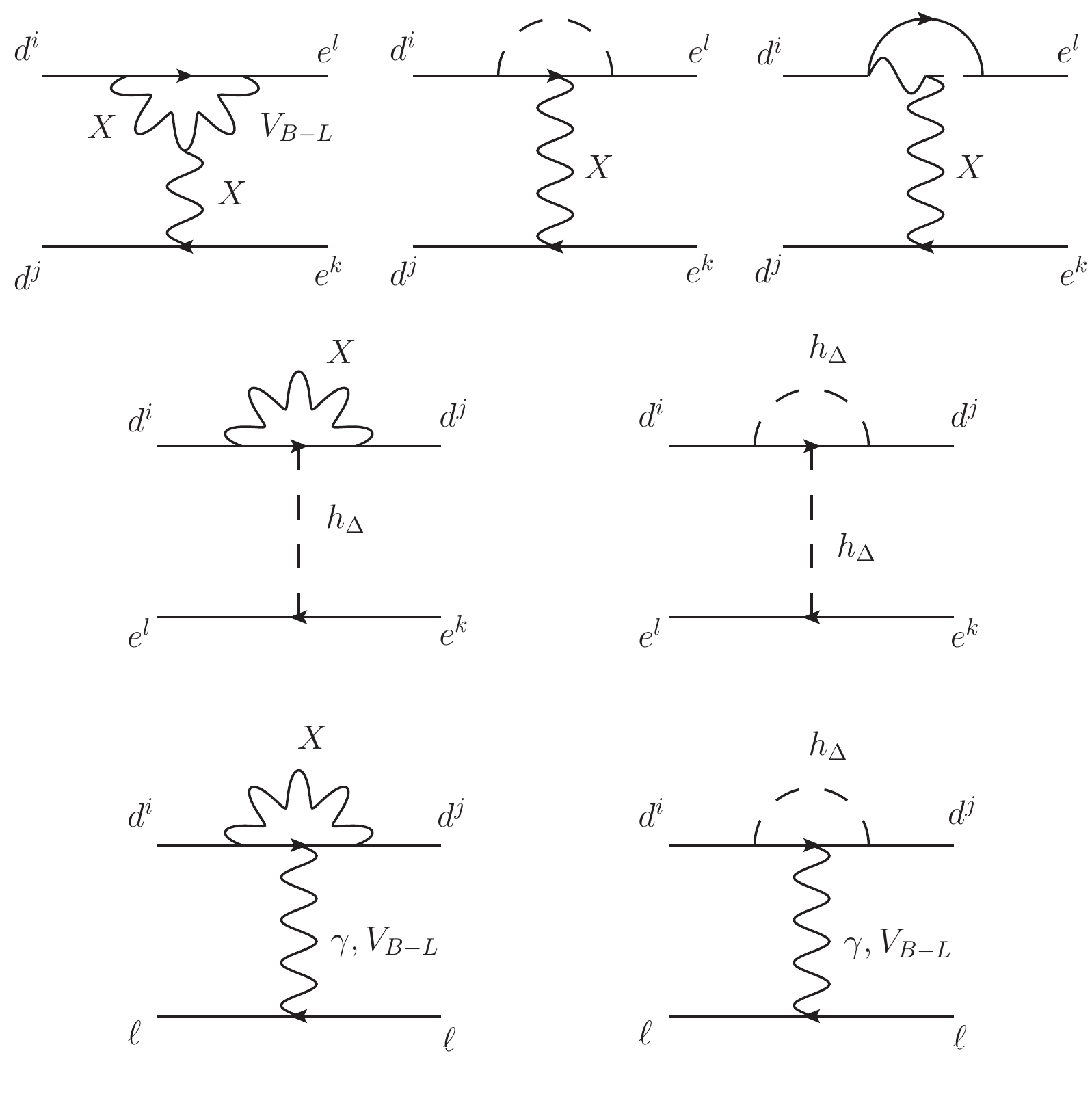}}
\caption{\label{penguin}
Example one-loop penguin contributions to semi-leptonic operators, $(\ol{d}_j \Gamma d_i) (\ol{e}_k \Gamma e_l)$ and $(\ol{d}_j \Gamma d_i) (\ol{\ell} \Gamma \ell)$.}
\end{figure}

We comment on contributions from one-loop penguin diagrams, 
which may generate the flavor violation comparable with the box contributions.\footnote{There are two-loop contributions: two sets of one-loop vertex corrections involving a LQ and an adjoint scalar.
However, such contributions are suppressed by the loop factor, couplings and the power of $m_{\VL}^2/m_{X}^2$ and thus expected to be smaller than the one-loop box contribution.}
Figure\,\ref{penguin} shows examples of the contributing diagrams. 
They fall into two categories; 
one violates both quark and lepton flavor and the other violates either of them but not both. 
The diagrams in the first and second lines of Fig.\,\ref{penguin} belong to the first category and 
can generate the quark and lepton flavor violating operators in the form $(\ol{d_j} \Gamma d_i)(\ol{e_k} \Gamma e_l)$, where $\Gamma$ represents an arbitrary Lorentz structure.
As discussed above, such operators lead to the $K_L \to \mu e$ decay and thus to the severe constraint.
These contributions, however, contain a tree-level coupling, namely $\Omega_{L,R}^{jk}$, $(\hat{Y}_\Delta^e)^{kl}$ or $(\hat{Y}_\Delta^d)^{ji}$, and are always suppressed under the condition (i). 

In the second category, the lepton flavor conserving processes $\ol{d_j} d_i \to \ell \bar{\ell}$ are induced via photon and $V_{B-L}$ penguin diagrams (the third line of Fig.\,\ref{penguin}) even though we require the condition (i). 
Here, let us have a closer look at the coupling structure of such contributions.
In the diagrams with the LQ loop, the amplitudes of such processes contain the following structures
\begin{equation}
\sum_I (\Omega_{L,R})^{jI} (\Omega_{L,R}^\dagger)^{Ii} \ f(m_I^e, m_X) , \quad 
\sum_I (\Omega_L)^{jI} (\Omega_R^\dagger)^{Ii} \ g(m_I^e, m_X) ,
\end{equation}
where $f$ and $g$ are loop functions. 
They depend on the LQ couplings to the vector-like families.
Assuming the condition (ii), the former contribution is vanishing for $i \neq j$ because of the unitarity of the LQ couplings $\Omega_{L,R}$, Eq.\,(\ref{eq-unitV3A}). 
The latter contribution is proportional to $(\Omega_L)^{jI} (\Omega_R^\dagger)^{Ii} \propto (\hat{Y}_\Delta^d)^{ji} + {\cal O}(\eta)$, and it is also vanishing. 
In the case of the $h_\Delta$ and $\Delta_8$ loops, the amplitudes contain the following coupling structures
\begin{equation}
\sum_I (\hat{Y}_\Delta^d)^{jI} (\hat{Y}_\Delta^{d\,\dagger})^{Ii} \ f'(m_I^d,m_{h_\Delta}) , \quad
\sum_I (\hat{Y}_\Delta^d)^{jI} (\hat{Y}_\Delta^d)^{Ii} \ g'(m_I^d,m_{h_\Delta}) ,
\end{equation}
where $f'$ and $g'$ are different loop functions from $f$ and $g$.
Under the conditions (i) and (ii), 
these are proportional to $\sum_I (\hat{Y}_\Delta^d)^{jI} (\hat{Y}_\Delta^{d\,\dagger})^{Ii} \simeq \delta^{ji}$ and 
$\sum_I (\hat{Y}_\Delta^d)^{jI} (\hat{Y}_\Delta^{d})^{Ii} \propto (\ol{\Psi}_L)^{ji} \simeq 0$, respectively, 
and thus the $\ol{d_j} d_i \to \ell \bar{\ell}$ processes do not appear. 
Moreover, other diagrams belonging to the second category can also induce the lepton flavor violating processes, such as $\mu \to e$ conversion and $\mu \to e \gamma$, but one can readily show that these are suppressed in a similar manner.
We thus conclude that the conditions (i) and (ii) are sufficient to suppress all penguin contributions in the model.

\section{Summary}
\label{summary}
In this paper, 
we study the one-loop contributions to the flavor violating processes, especially  
the LFV meson decays and $\mu\to e$ conversion, in the PS model with vector-like families.
These processes are known to strongly constrain the scale of the PS symmetry breaking.
We clarify the conditions to suppress these processes up to at the one-loop level: 
\begin{enumerate}[(i)]
\item {LQ couplings to the SM families are vanishing, i.e. $X_L = X_R \sim 0$,} 
\item {Masses of vector-like down-type quarks and charged leptons are individually universal,}
\item {$\Ups_L$ and $\Ups_R$ have a certain structure such that 
$K_L\to\mu e$ and $\mu\to e$ conversion are sufficiently small.} 
\end{enumerate}
These are the conditions at the leading order in $\eta := m_{\alpha \beta}/v_\Delta$.

The condition (i) is required to suppress the flavor violating processes mediated by the tree-level LQ exchange, while 
the one-loop box diagrams with two LQs can induce those processes only with the condition (i).
Note that the tree-level flavor violations via the scalar fields 
are suppressed independently of the condition (i), as shown in Eqs.\,(\ref{eq:Omega_LDR}) and (\ref{eq:Omega_LER}). 
Once we impose the condition (ii) as well as the condition (i),
the flavor violating processes from the box diagrams with two LQs are suppressed due to the unitarity of the LQ gauge couplings, in analogy with the $W$ boson coupling in the SM. 
We also argued that four-quarks, four-leptons and penguin operators are suppressed due to the unitarity. 
Therefore, the flavor violating processes like neutral meson mixing, 
$\mu\to e\gamma$ and $\mu \to eee$ are all suppressed under the conditions (i) and (ii).

Nonetheless, the condition (iii) is necessary to alleviate the constraints from the LFV processes induced by the box diagrams involving both vector LQ and scalars. We found that those diagrams are not suppressed even with the conditions (i) and (ii). 
Such contribution is well represented by a coupling structure $\Ups_A$ ($A=L,R$) defined in Eq.\,\eqref{eq-Ups}, whose SM blocks $\Ups_A^{ij}$ are the $3\times3$ unitary matrices.  
Because of the unitarity of $\Ups_A^{ij}$, we cannot realize $\Ups_A^{ij} = 0$ and 
hence the flavor violation via this coupling structure is unavoidable. 
As a result, $\Ups_A^{ij}$ should have a structure 
that sufficiently suppresses the LFV processes, especially $K_L \to \mu e$ and $\mu\to e$ conversion. 
This is what the condition (iii) means.

To evaluate the change of the limits depending on the $\Ups_A^{ij}$ structure, 
we studied three simplified cases given in  Eq.\,\eqref{eq-caseAB}.
In the case (a), the structure of $\Ups_A^{ij}$ allows 
the chiral enhanced $K_L \to \mu e$ decay,
and hence $\order{10~\TeV}$ LQ mass is required to be consistent 
with the experimental limit. 
It is remarkable that there are upper bounds on the vector-like fermion masses 
of $\order{\TeV}$ to respect $K_L \to \mu e$. 
This result would encourage direct searches for the vector-like fermions 
at the LHC and other future collider experiments to test this scenario. 
In the case (b), 
$K_L \to \mu e$ is not induced because of the structure of $\Ups_A^{ij}$,
but $\mu\to e$ conversion arises. 
The resulting lower limit on the LQ mass is 6 TeV currently 
and it will be improved to 15 TeV at the future experiment with the aluminum target. 
In the case (c), both $K_L\to \mu e$ and $\mu \to e $ conversion are absent, 
so $B_d \to \mu e$ gives the strongest limit. 
Since the experimental limits are much weaker for this decay mode, 
$5~\TeV$ LQ is not excluded.

In conclusion, 
while we suggested the conditions (i)-(iii) to suppress the flavor violating processes,
it will be interesting to study what will be caused by the violation of these conditions. 
In particular, 
the tree-level contributions, namely the violation of the condition (i) 
is required to address the $R_{K^{(*)}}$ anomaly, 
so the full numerical analysis with both tree-level and one-loop contributions 
is crucial. 
Furthermore, this model could perhaps explain the anomaly in muon $g-2$ 
via the loop diagrams involving LQ, vector-like fermions and exotic scalar particles. 
We leave those extended studies including the violation of three conditions for future work.

\section*{Acknowledgment}
S. I. enjoys the support from the Japan Society for the Promotion of Science (JSPS) Core-to-Core Program, No.JPJSCCA20200002 and the Deutsche Forschungsgemeinschaft (DFG, German Research Foundation) under grant 396021762–TRR\,257.
The work of J.K.
is supported in part by
the Institute for Basic Science (IBS-R018-D1)
and the Grant-in-Aid for Scientific Research from the
Ministry of Education, Science, Sports and Culture (MEXT), Japan No.\ 18K13534.
S.O. acknowledges financial support from the State Agency for Research of the Spanish Ministry of Science and Innovation through the ``Unit of Excellence Mar\'ia de Maeztu 2020-2023'' award to the Institute of Cosmos Sciences (CEX2019-000918-M) and from PID2019-105614GB-C21 and 2017-SGR-929 grants.
The work of Y. O. is supported by Grant-in-Aid for Scientific research from the MEXT, Japan, No. 19K03867.

\appendix
\section{Loop functions}
\label{app-Int}
The loop functions are shown in this appendix. 
First, we define
\begin{align}
F_n(m_1,m_2;M_1,M_2):= 
 \int \frac{d^4 p}{(4 \pi)^4} 
 \frac{ p^{2n}}{ (p^2-m^2_1)(p^2-m^2_2)(p^2-M^2_1)(p^2-M^2_2)} .
\end{align}
The functions with $n=0,1$ are relevant to our study, which are given by 
\begin{align}
F_0(m_1,m_2;M_1,M_2)&=\  \frac{-i}{16 \pi^2} \frac{1}{M^4_1}  \left \{ F(x_1, x_2, \eta) + F(x_2, x_1, \eta) + F(\eta,x_1, x_2)  \right \},  \\
F_1(m_1,m_2;M_1,M_2)&=\ \frac{-i}{16 \pi^2} \frac{1}{M^2_1}  \left \{x_1 F(x_1, x_2, \eta) +x_2 F(x_2, x_1, \eta) + \eta F(\eta,x_1, x_2)  \right \},
\end{align}
where $x_1={m^2_1}/{M^2_1}$, $x_2={m^2_2}/{M^2_1}$, $\eta={M^2_2}/{M^2_1}$. 
The function $F$ is defined as
\begin{align}
F(x_1,x_2,\eta)= \frac{x_1 \ln x_1}{(x_1-1)(x_1-x_2)(x_1-\eta)}.
\end{align} 

In the simplified analysis in Sec.~\ref{sec-pheno}, we consider the case of $m:= m_1 = m_2$ and $M:= M_1 = M_2$. 
In this case, $F_0$ and $F_1$ take the simplified forms, 
\begin{align}
 F_0(m,M) =&\ \frac{-i}{16\pi^2 M^4}  \frac{2(1-x)+(1+x) \log x}{(1-x)^3}, \label{eq:loopfunc_approx1}  \\ 
 F_1(m,M) =&\ \frac{-i}{16\pi^2 M^2}  \frac{1-x^2+2x\log x}{(1-x)^3},  \label{eq:loopfunc_approx2}
\end{align}
where $x:=m^2/M^2$.
The dimensionless functions $G_n$ are defined as the linear combinations of $F_0$ and $F_1$,
\begin{align}
m_X^{2n-4}G_n(m_E,m_D;m_X,m_{h_\Delta}):=& \ 
 \frac{3}{8} F_{n}(m_E,m_E;m_X,m_{h_\Delta})
+\frac{11}{8} F_{n}(m_D,m_D;m_X,m_{h_\Delta})  \notag \\ 
&+ \frac{1}{4}F_{n}(m_E,m_D;m_X,m_{h_\Delta}).
\end{align}
We also define 
$\tF_0(m_E, m_D; m_X, m_{h_\Delta}):= m_X^4 F_0(m_E, m_D; m_X, m_{h_\Delta})$, 
where $\tF_0$ is a dimensionless function. 

\section{Tree-level constraints}
\label{sec:tree_constraint}

The tree-level LQ exchange induces the LFV processes. 
In particular, $e\mathrm{-}\mu$ flavor violating phenomena strongly constrain the coupling products involving the first two generations.
The prime constraints arise from the LFV meson decays, especially $K_L\to \mu e$.
The $\mu\mathrm{-}e$ conversion process also brings a stringent constraint on the quark-flavor-diagonal coupling products.
In order to avoid the experimental constraints, those couplings need to be smaller than $\mathcal{O}(10^{-2\sim3})$, assuming that all the couplings have a comparable size. 
See Table\,\ref{Tab:LQcoup}, where we fix $m_X=5\,\TeV$. 
Furthermore one-loop induced $\mu\to e\gamma $ can constrain the coupling products. It however depends on the LQ couplings to the vector-like quarks \cite{Iguro:2021kdw}, so we do not discuss it further.
For the more generic analysis readers are referred to Ref.\cite{Fornal:2018dqn}. 

\begin{table}[h]
\centering
 \caption{\label{Tab:LQcoup} 
    The upper limit on the coupling products from the $\mu\mathrm{-}e$ flavor violating processes.
}
\newcommand{\bhline}[1]{\noalign{\hrule height #1}}
\renewcommand{\arraystretch}{1.3}
   \scalebox{0.94}{
  \begin{tabular}{c|  c | c | c } \hline
  \rowcolor{white} 
  coupling product & upper limit & process& bound\\  \hline \hline 
 $\left | \left (\hat{g}^X_{d_L} \right )_{d e }\left (\hat{g}^X_{d_R} \right )_{s \mu } \right |, \left | \left (\hat{g}^X_{d_R} \right )_{d e }\left (\hat{g}^X_{d_L} \right )_{s \mu } \right |$ & $\lesssim 10^{-5}$ &  $K_L\to\mu e$ & \cite{Zyla:2020zbs}\\  
 $\left | \left (\hat{g}^X_{d_L} \right )_{s e }\left (\hat{g}^X_{d_R} \right )_{d \mu } \right |, \left | \left (\hat{g}^X_{d_R} \right )_{s e }\left (\hat{g}^X_{d_L} \right )_{d \mu } \right |$ & $\lesssim 10^{-5}$ & $K_L\to\mu e$ & \cite{Zyla:2020zbs}\\  \hline 
  $\left | \left (\hat{g}^X_{d_{L}} \right )_{d e }\left (\hat{g}^X_{d_{L}} \right )_{d \mu } \right |$ & $\lesssim 10^{-4}$ & $\mu\mathrm{-}e$ conversion & \cite{Bertl:2006up}\\  
 $\left | \left (\hat{g}^X_{d_{R}} \right )_{d e }\left (\hat{g}^X_{d_{R}} \right )_{d \mu } \right |$  & $\lesssim 10^{-5}$ & $\mu\mathrm{-}e$ conversion & \cite{Bertl:2006up}\\ 
 $\left | \left (\hat{g}^X_{d_{L}} \right )_{d e }\left (\hat{g}^X_{d_{R}} \right )_{d \mu } \right |$ & $\lesssim 10^{-5}$ & $\mu\mathrm{-}e$ conversion & \cite{Bertl:2006up}\\ 
 $\left | \left (\hat{g}^X_{d_{L}} \right )_{s e }\left (\hat{g}^X_{d_{R}} \right )_{s \mu } \right |$ & $\lesssim 10^{-5}$ & $\mu\mathrm{-}e$ conversion & \cite{Bertl:2006up} \\  \hline 
   \end{tabular}
   }
\end{table}

The $h_{\Delta}$ and $\Deladj$ exchanging also contributes to the flavor violating processes.
The scalar couplings with the SM fermions are generally flavor violating and
linear to
\beq
 \left(\hat{g}^X_{d_{L}} \right)_{iD} m^e_{D}\left(\hat{g}^X_{d_{R}} \right)^\dagger_{Dj},~\left(\hat{g}^X_{d_{L}} \right)^\dagger_{iD} m^d_{D}\left(\hat{g}^X_{d_{R}} \right)_{Dj}.
\eeq
As we see in the main text, the scalar couplings are related to the LQ couplings involving heavy fermions as well as SM fermions.
Taking into account the tree-level exchanging of  $h_{\Delta}$ and $\Deladj$ and fixing $m_{h_\Delta}=m_{\Delta_8}=5$\,TeV, we can also derive the experimental constraints on the scalar couplings
\begin{eqnarray}
\frac{1}{3}\sqrt{\frac{3}{8}}\left ( \left | \Hat Y^d_{\Delta \, ij } \right | \right ) &\leq&  
\begin{pmatrix} 5 \times 10^{-3} & 10^{-5} & 10^{-3} \\  10^{-5} & 1 \times 10^{-2} & 5 \times 10^{-3} \\ 10^{-3} &  5 \times 10^{-3} & 1  \end{pmatrix},  \\
\sqrt{\frac{3}{8}}\left ( \left | \Hat Y^e_{\Delta \, ij } \right | \right )  & \leq & 
\begin{pmatrix} 1 & 1 \times 10^{-3} & 0.5 \\  1 \times 10^{-3} & 1 & 0.5 \\ 0.5 &  0.5 & 1  \end{pmatrix}.
\end{eqnarray}
The upper bounds on the off-diagonal elements of $(\Hat Y^d_{\Delta})_{ij}$ are estimated using $K$-$\ov{K}$, $B$-$\ov{B}$ and $B_s$-$\ov{B_s}$ mixings. 
The bounds on the off-diagonal elements of $(\Hat Y^e_{\Delta})_{ij}$ are derived from $\mu \to 3e$, $\tau \to \ell \ell \ell^{\prime}$, by setting all diagonal elements to unity.
The bounds on the diagonal elements of $(\Hat Y^d_{\Delta})_{ij}$ are derived from the $\mu$-$e$ conversion with the maximally allowed $(\Hat Y^e_{\Delta})_{e\mu}$.
As discussed in the main text, the conditions (i) and (ii) suppress those dangerous couplings.


\bibliographystyle{JHEP}
\bibliography{PSleptoquark} 

\end{document}